\documentclass{jfm}

\usepackage{graphicx}
\usepackage{natbib}

\ifCUPmtlplainloaded \else
  \checkfont{eurm10}
  \iffontfound
    \IfFileExists{upmath.sty}
      {\typeout{^^JFound AMS Euler Roman fonts on the system,
                   using the 'upmath' package.^^J}%
       \usepackage{upmath}}
      {\typeout{^^JFound AMS Euler Roman fonts on the system, but you
                   dont seem to have the}%
       \typeout{'upmath' package installed. JFM.cls can take advantage
                 of these fonts,^^Jif you use 'upmath' package.^^J}%
      }
  \else
  \fi
\fi

\ifCUPmtlplainloaded \else
  \checkfont{msam10}
  \iffontfound
    \IfFileExists{amssymb.sty}
      {\typeout{^^JFound AMS Symbol fonts on the system, using the
                'amssymb' package.^^J}%
       \usepackage{amssymb}%
       \let\le=\leqslant  
         
      }{}
  \fi
\fi

\ifCUPmtlplainloaded \else
  \IfFileExists{amsbsy.sty}
    {\typeout{^^JFound the 'amsbsy' package on the system, using it.^^J}%
     \usepackage{amsbsy}}
    {}
\fi

\newsavebox{\astrutbox}
\sbox{\astrutbox}{\rule[-5pt]{0pt}{20pt}}

\title[Reduced description of exact coherent states in parallel shear flows]{Reduced description of exact coherent states in parallel shear flows}

\author[C. Beaume, G. P. Chini, K. Julien and E. Knobloch]%
{C\ls \'E\ls D\ls R\ls I\ls C\ns B\ls E\ls A\ls U\ls M\ls E$^1$%
  \thanks{Email address for correspondence: ced.beaume@gmail.com},\ns
G\ls R\ls E\ls G\ls O\ls R\ls Y\ns P.\ns C\ls H\ls I\ls N\ls I$^2$\thanks{Email address for correspondence: greg.chini@unh.edu} \ns
K\ls E\ls I\ls T\ls H\ns J\ls U\ls L\ls I\ls E\ls N$^3$\thanks{Email address for correspondence: keith.julien@colorado.edu}
\and 
E\ls D\ls G\ls A\ls R\ns K\ls N\ls O\ls B\ls L\ls O\ls C\ls H$^1$\thanks{Email address for correspondence: knobloch@berkeley.edu}}

\affiliation{$^1$Department of Physics, University of California, Berkeley CA 94720, USA;\\[\affilskip]
$^2$Department of Mechanical Engineering \& Program in Integrated Applied Mathematics, University of New Hampshire, Durham NH 03824;\\[\affilskip]
$^3$Department of Applied Mathematics, University of Colorado at Boulder, Boulder CO 80309}

\pubyear{2010}
\volume{650}
\pagerange{119--126}
\date{?; revised ?; accepted ?. - To be entered by editorial office}
\begin{document}

\maketitle

\begin{abstract}
Exact coherent states of a linearly stable, plane parallel shear flow confined between stationary stress-free walls and driven by a sinusoidal body force (a flow first introduced by F. Waleffe, Phys. Fluids {\bf 9}, 883 (1997)) are computed using equations obtained from a large Reynolds-number asymptotic reduction of the Navier-Stokes equations. The reduced equations employ a decomposition into streamwise-averaged (mean) and streamwise-varying (fluctuation) components and are characterized by an effective order one Reynolds number in the mean equations along with a formally higher-order diffusive regularization of the fluctuation equations. A robust numerical algorithm for computing exact coherent states is introduced.  Numerical continuation of the lower branch states to lower Reynolds numbers reveals the presence of a saddle-node; the saddle-node allows access to upper branch states that, like the lower branch states, appear to be self-consistently described by the reduced equations. Both lower and upper branch states are characterized in detail.

\end{abstract}

\begin{keywords}
Authors should not enter keywords on the manuscript, as these must be chosen by the author during the online submission process and will then be added during the typesetting process (see http://journals.cambridge.org/data/\linebreak[3]relatedlink/jfm-\linebreak[3]keywords.pdf for the full list)
\end{keywords}

\section{Introduction}\label{INTRO}

Exact, fully nonlinear, three-dimensional (3D) solutions of the Navier-Stokes equations play an important role in our understanding of the transition to turbulence in parallel shear flows and of the recurrence properties of the turbulence that results. These solutions, first computed by \cite{Nagata90} and \cite{Waleffe97} and now called \emph{exact coherent states} (ECS), may take the form of time-independent states (i.e., equilibria), or time-periodic states (e.g., traveling waves). ECS have now been computed by numerous investigators for a number of different flows, including plane Couette flow \citep{Gibson08} and pipe flow \citep{Duguet08,Duguet10}. Typically, these solutions consist of streamwise-oriented streaks and vortices that bear a striking qualitative and even quantitative resemblance to the coherent structures commonly observed in turbulent wall flows, although they are generally unstable. In fact, despite their instability, ECS are frequently observed as transients in both shear flow simulations and experiments. Analysis of a low-order model by \cite{Waleffe97} and a more systematic numerical study by \cite{Schmiegel99} reveal that the ECS in plane Couette flow (PCF) are born in a saddle-node bifurcation as the Reynolds number $Re$ increases, and continue as upper- and lower-branch solutions; this ECS bifurcation scenario seems generic in that it is commonly found in other shear flows.  Much of the interest in ECS can be attributed to the possibility that upper-branch solutions comprise the ``skeleton" of a high-dimensional ``turbulent'' attractor in the shear-flow phase space. Both \cite{Waleffe01} and \cite{Kawahara01} demonstrate that certain low-order statistics of PCF turbulence, particularly the mean and root-mean-square (rms) velocity profiles, can be accurately reproduced using these unstable solutions, quantitatively attesting to their physical relevance.

In two particularly insightful papers, \cite{Waleffe97,Waleffe01} identifies a fully nonlinear process, the self-sustaining process, involving the interaction of streamwise-oriented streaks and rolls that sustains lower-branch ECS in plane parallel shear flows (including PCF). Certain of these solutions have the remarkable property that they have only a \emph{single} unstable eigendirection. These lower-branch coherent states appear to separate, in phase space, disturbances that decay, causing relaminarization of the flow, from those that follow an excursion toward a turbulent (or at least transiently chaotic) state.  For this reason, investigation of these ``edge states" and their stability is of great interest, offering tantalizing opportunities for flow control. Importantly, the recent discovery by \cite{Schneider10} of spatially localized edge states in PCF has served to further increase interest in ECS by establishing that they are not special solutions found only in small laterally-periodic domains but, rather, that they exist in unbounded flows and may thus play a role in transition in open shear flows. 

The focus of the present work is on a family of \emph{equilibrium} ECS in a body-forced parallel shear flow first proposed by \cite{Waleffe97}, which we refer to as ``Waleffe flow" (hereafter WF, see \S\ref{PDEs}). This flow is a close relative of PCF, implying a likely connection between the ECS we compute and those found by \cite{Nagata90}, \cite{Clever97} and later continued via homotopy by \cite{Waleffe03} to other flows.  Indeed, our formulation is readily adapted to treat a variety of plane parallel shear flows, including PCF and plane Poiseuille flow. In general, the extraction of both lower and upper branch ECS in PCF, as in other shear flows, has required substantial computational effort.  In particular, because ECS are typically disconnected from the structureless base shear flow, standard methods based on linear theory cannot be used to identify them.  Instead, more sophisticated algorithms (or considerable ingenuity and physical intuition -- see \cite{Waleffe97, Waleffe01}) are required. These challenges are exacerbated for flows at large Reynolds numbers and in large domains.

For these reasons semi-analytical approaches to the problem of finding ECS and more generally edge states are invaluable. A significant advance in this direction has recently been achieved through the pioneering work of \cite{Hall10} and subsequently \cite{Blackburn13}, who describe a procedure for computing the Nagata--Clever/Busse--Waleffe lower branch ECS for PCF in the asymptotic limit of large Reynolds numbers, i.e., precisely the regime that is inaccessible to strictly numerical approaches. These authors decompose the flow into streamwise-invariant (i.e., mean) and streamwise-varying (i.e., fluctuation, or ``wave") components and exploit the remarkable scaling properties of these fields.  These properties, first introduced by \cite{Hall91b} and empirically observed by \cite{Wang07}, establish that the lower branch edge states have an asymptotic structure, as $Re\to\infty$, consisting of $\mathit{O}(1)$ streamwise-invariant streaks and $\mathit{O}(Re^{-1})$ streamwise-invariant rolls. Fundamental streamwise-varying modes also scale roughly as $\mathit{O}(Re^{-1})$ -- \cite{Wang07} cite an exponent of approximately $-0.9$ -- but higher harmonics are found to be $\mathit{o}(Re^{-1})$, see table \ref{scaling}. 
\begin{table}
\begin{center}
\def~{\hphantom{0}}
    \begin{tabular}{ccccccc}
    ~Mode~ & ~$u_0$~ & ~$(v_0,w_0)$~ & ~$(u_1,v_1,w_1)$~ & ~$(u_2,v_2,w_2)$~ & ~$(u_3,v_3,w_3)$~ & ~$(u_n,v_n,w_n)$~ \\ [3pt]
    ~Scaling~ & ~$\mathit{O}(1)$~ & ~$\mathit{O}(Re^{-1})$~ & ~$\mathit{O}(Re^{-0.9})$~ & ~$\mathit{O}(Re^{-1.6})$~ & ~$\mathit{O}(Re^{-2.2})$~ & ~$o(Re^{-2.2})$~ \\
    \end{tabular}
    \caption{Summary of the scalings obtained by \cite{Wang07} for lower branch ECS in plane Couette flow.  These authors decompose the ECS into streamwise Fourier modes:  ${\bf u} (x,y,z,t) = y {\bf \hat{x}} + \sum_{n=-N/2}^{N/2} {\bf u_n}(y,z) e^{in\theta} + c.c.$, where $n$ denotes the index of the Fourier mode ${\bf u_n}$, $N$ is the number of Fourier modes retained and $\theta = \alpha (x-ct)$. Here $\alpha$ is the fundamental streamwise wavenumber, $c$ is the speed of the wave (in the case of a traveling wave solution), and $c.c.$ denotes the complex conjugate. The last column shows that higher harmonics ($n>3$) decay faster than the primary ones.}
\label{scaling}
\end{center}
\end{table}
In this way, \cite{Hall10} are able to reduce the computation of these ECS to the solution of a two-dimensional (2D) system for the streamwise-averaged fields at \emph{unit} rescaled Reynolds number coupled to a quasi-linear inviscid eigenvalue problem for \emph{neutral} disturbances to a mean streaky streamwise ($x$-directed) flow $u_0(y,z)$, where $y$ and $z$ are wall-normal and spanwise coordinates, respectively. This eigenvalue problem is singular, possessing a non-planar critical layer at $u_0(y,z)=0$. The authors perform a careful matched asymptotic analysis to incorporate a viscously regularized critical layer, ultimately deriving jump conditions across the layer that link the mean fields on either side.  In addition to reducing the computational cost of numerically solving for the lower-branch ECS at large $Re$, their analysis clearly demonstrates the physical mechanism by which the fluctuations sustain the mean fields:  namely, steady streaming is driven within the critical layer, which in turn drives the mean flow outside the layer.

For all its merit, the approach of \cite{Hall10} is, for the uninitiated, rather formidable. Furthermore, despite its elegance, the mathematical model they derive requires two forms of \emph{regularization} to render it suitable for numerical computation of ECS.  First, the inviscid eigenvalue problem (a generalized, two-dimensional version of Rayleigh's equation first obtained by \cite{Hall91}) is regularized via the introduction of a pseudo-Reynolds number. Secondly, the jump conditions are enforced by introducing in the mean equations a delta-function-like body force, which must be suitably smoothed.  Moreover, \cite{Hall10} employ a sophisticated high-order domain-decomposition numerical scheme, which requires the numerical grid to be adaptively updated since the location of the critical layer is not known \emph{a priori}.  


To overcome these difficulties, \cite{Blackburn13} proposed a hybrid asymptotic--computational approach that avoids the intricacies of imposing and then regularizing jump conditions across the critical layer while still leveraging the large-$Re$ asymptotic reduction.  Our approach, which likewise exploits the large-$Re$ scalings first reported by \cite{Wang07} for lower-branch ECS in PCF, is closely related to the formulation of \cite{Blackburn13} but was developed independently \citep{BeaumeGFD12} using an asymptotic reduction methodology originally introduced for high Reynolds-number flows subjected to strong constraints \citep{JK07, CJK09}. The present study extends this work in several important ways. First, as noted above, we treat WF rather than PCF, for which the detailed structure of the ECS necessarily differs. Secondly, we use asymptotic analysis to motivate a \emph{composite} multiscale PDE model \citep{Giannetti06}, in which formally small diffusion terms are retained because they assume leading-order importance in thin critical and boundary layers. This derivation, which is given in \S\ref{PDEs}, highlights the underlying partial differential equation (PDE) structure associated with the formation of ECS and also reveals how slow streamwise modulation of the mean and fluctuation fields may be consistently incorporated.  Crucially, the resulting wave/mean-flow equations are uniformly valid over the entire domain, obviating the need for explicit introduction and subsequent smoothing of jump conditions and for any further regularization of the fluctuation equations. Since jump conditions are not imposed, there is no need for adaptive mesh refinement associated with dynamic tracking of the critical layer. Of course, for very large $Re$ sufficiently many modes or grid points must be used to resolve the inevitable sharp gradient regions. In view of this restriction, one advantage of investigating Waleffe flow instead of plane Couette flow is that trigonometric basis functions may be employed in \emph{both} the wall-normal and spanwise directions to provide a higher mesh density within the critical layer (\S\ref{NUMERICS}); that is, we employ -- and, as necessary, refine -- an \emph{equispaced} grid in both coordinate directions.  In contrast, Chebyshev polynomials, commonly used in spectral simulations of PCF, would yield least resolution where it is most desired: in the neighborhood of the critical layer. Finally, and somewhat remarkably, we demonstrate in \S\ref{RESULTS} that our asymptotically-reduced PDE model admits both lower branch and \emph{upper branch} solutions: in spite of the large Reynolds number formulation, the asymptotics prove sufficiently robust to capture the saddle node bifurcation at which both the lower and upper branch ECS are born.  Thus, our asymptotically-reduced PDEs should prove useful for a variety of further studies of parallel shear flows that aim, for example, to investigate streamwise and spanwise localization.



\section{Multiscale system}\label{PDEs}

\begin{figure}
\centerline{\includegraphics[width=\textwidth]{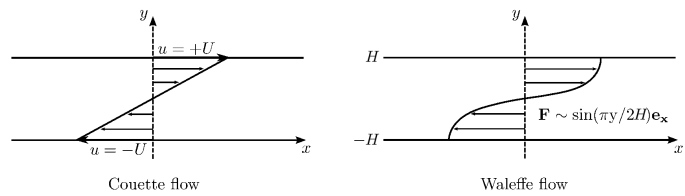}}
\caption{Sketch of plane Couette and Waleffe flows. Plane Couette flow is driven by wall motion in the $x$ direction, the top and bottom walls moving with opposite velocities $\pm U$. Plane Waleffe flow is driven by an $x$-directed body force. The forcing profile is a half-period of a sinusoid in $y$. Note that the laminar flow is stable despite the presence of an inflection point owing to the proximity of the (stress-free) walls.}
\label{couleffe}
\end{figure}

Incompressible channel flow driven by a volume force $\mathbf{f}(y)$ is governed by the nondimensional Navier--Stokes equations
\begin{eqnarray}
\partial_t \mathbf{v} + (\mathbf{v} \cdot \nabla) \mathbf{v}&=&-\nabla p\,+\,\frac{1}{Re}\nabla^2\mathbf{v}\,+\mathbf{f}(y),
\label{NSeqn}
\end{eqnarray}
along with the incompressibility constraint
\begin{eqnarray}
\nabla\cdot\mathbf{v}&=&0.\label{CONT}
\end{eqnarray}
Here and throughout, a Cartesian coordinate system is adopted in which $x$, $y$ and $z$ are the dimensionless streamwise, wall-normal and spanwise directions, respectively. The velocity vector $\mathbf{v}$ has dimensionless components $(u,v,w)$, and $p$ is the dimensionless fluid pressure. In (\ref{NSeqn}) all lengths have been scaled by $H$, i.e. half the (dimensional) distance separating the plane parallel walls, and all velocities by a characteristic velocity $U$.  In PCF, $U$ is the dimensional speed of the upper wall, the flow being driven by in-plane but opposing motion of the no-slip boundaries; in this case the body force vanishes: $\mathbf{f}(y)=\mathbf{0}$.  This configuration admits a structureless laminar solution, namely Couette flow, as depicted in figure~\ref{couleffe} (left panel).  As is well known, this solution is linearly stable even for asymptotically large values of the Reynolds number $Re\equiv UH/\nu$ \citep{Romanov73,Schmid01}, although stability is observed experimentally only for $Re<Re_u \approx 310$ \citep{Dauchot95a,Dauchot95b,Tillmark95}. Careful parameter studies have revealed, for larger Reynolds numbers, a variety of structured flow regimes \citep{Manneville04}: in $Re_u<Re<Re_g \approx 325$, perturbations to Couette flow evolve into evanescent turbulent spots before Couette flow is restored.  The lifetime of the transient spots diverges as $Re_g$ is approached \citep{Bottin98}, indicating the onset of sustained turbulence above $Re_g$, where most of the turbulent spots survive and organize themselves into turbulent bands oblique to the streamwise direction \citep{Prigent02,Barkley05}.  As $Re$ is increased further, turbulence progressively invades the domain until $R_t \approx 415$, where space-filling turbulence is observed.  

We focus here on a close relative of PCF, namely Waleffe flow (WF), depicted in figure~\ref{couleffe} (right panel). This flow is driven by an $x$-directed body force that varies sinusoidally in the wall-normal direction, \emph{viz.}, $\mathbf{f}(y)=\frac{\sqrt{2}\pi^2}{4Re}\sin\left(\frac{\pi y}{2}\right)\mathbf{\hat{x}}$, where $\mathbf{\hat{x}}$ is a unit vector in the $x$ direction.  Moreover, stress-free rather than no-slip conditions are imposed along stationary boundaries located at $y=\pm1$.  This flow was suggested by \cite{Waleffe97} as an alternative to PCF that is more convenient for low-order modeling.  Indeed, WF can be naturally expanded in Fourier modes in all three coordinate directions, and the laminar basic state $\mathbf{v}$=($\sqrt{2}\sin(\pi y/2)$,0,0) is itself a low-order mode in this basis.  Note that in WF the velocity scale $U$ is the root-mean-square velocity of the corresponding dimensional laminar base flow.  Although this base flow has an inflection point, it is nevertheless linearly stable for all $Re$ \citep{Drazin81}. This is a consequence of the blocking effect of the walls at $y=\pm1$; in contrast, the related Kolmogorov flow \citep{Arnold60} is defined with periodic boundary conditions in $y$, thereby eliminating the stabilizing effect of the walls and permitting linear instability on large scales \citep{Meshalkin61,Love99,Lucas13}. However, WF does admit finite amplitude solutions supported by the self-sustaining process identified by \cite{Waleffe97}. These ECS also cannot bifurcate from the base flow, but instead appear through saddle-node bifurcations as $Re$ increases, much as in PCF. 

\subsection{Multiscale asymptotic analysis}

In this section we derive the basic equations used in this paper. The procedure is motivated by earlier work on flows with strong restraints \citep{JK07}, particularly Langmuir circulation \citep{CJK09}. In these flows the strong restraining force reduces the effective dimensionality of the system, leading to a simplified description. 

We begin by decomposing the velocity $\mathbf{v}$ into a streamwise component $u$ and the perpendicular components $\mathbf{v}_\bot=(v,w)$, i.e., $\mathbf{v}\equiv(u,\mathbf{v}_\bot)$. Thus
\begin{eqnarray}
\partial_t u+u\partial_x u+\left(\mathbf{v}_\bot\cdot\nabla_\bot\right)u
   &=&-\partial_x p + \frac{1}{Re}\left(\partial_x^2+\nabla_\bot^2\right)u + \frac{\sqrt{2}\pi^2}{4Re}\sin\left(\frac{\pi y}{2}\right),\label{xeqn}\\
\partial_t\mathbf{v}_\bot+u\partial_x\mathbf{v}_\bot
   +\left(\mathbf{v}_\bot\cdot\nabla_\bot\right)\mathbf{v}_\bot
   &=&-\nabla_\bot p + \frac{1}{Re}\left(\partial_x^2+\nabla_\bot^2\right)\mathbf{v}_\bot,
   \label{perpeqn}\\
\partial_x u + \nabla_\bot\cdot\mathbf{v}_\bot&=&0\label{continuity}.
\end{eqnarray}
These equations are supplemented with the following boundary conditions along the walls at $y=\pm1$: 
\begin{equation}
\partial_y u = v = \partial_y w = 0. \label{boundary}
\end{equation}
In addition, all fields are taken to be periodic in the streamwise and spanwise directions, with periods $L_x$ and $L_z$, respectively.

To allow for slow modulation in the streamwise direction, we explicitly introduce a slow streamwise coordinate $X\equiv\epsilon x$ and an associated slow time scale $T\equiv\epsilon t$, where the small parameter $\epsilon\equiv 1/Re$.  Thus, all field variables are interpreted as functions of $(x,X,y,z,t,T)$. Replacing the original $x$ and $t$ derivatives by $\partial_x+\epsilon\partial_X$ and $\partial_t+\epsilon\partial_T$, respectively, Eqns.~(\ref{xeqn})--(\ref{continuity}) become
\begin{eqnarray}
\left[\partial_t+\epsilon\partial_T\right]u+\left[\partial_x+\epsilon\partial_X\right]u^2
+\nabla_\bot\cdot\left(\mathbf{v}_\bot u\right)&=&-[\partial_x+\epsilon\partial_X]p+\epsilon\frac{\sqrt{2}\pi^2}{4}\sin\left(\frac{\pi y}{2}\right)\nonumber\\
&&+\epsilon\left[\partial_x^2+2\epsilon\partial_X\partial_x+\epsilon^2\partial_X^2+\nabla_\bot^2\right]u,\nonumber\\
\label{VxEQN}\\
\left[\partial_t+\epsilon\partial_T\right]\mathbf{v}_\bot+\left[\partial_x+\epsilon\partial_X\right]
\left(u\mathbf{v}_\bot\right)+\nabla_\bot\cdot\left(\mathbf{v}_\bot\mathbf{v}_\bot\right)
&=&-\nabla_\bot p\nonumber\\
&&+\epsilon\left[\partial_x^2+2\epsilon\partial_X\partial_x+\epsilon^2\partial_X^2+\nabla_\bot^2\right]
\mathbf{v}_\bot,\nonumber\\
\label{VpEQN}\\
\left[\partial_x+\epsilon\partial_X\right]u + \nabla_\bot\cdot\mathbf{v}_\bot&=&0.\label{CONTEQN}
\end{eqnarray}
Next, we decompose all variables into a fast $(x,t)$ average plus a fluctuation with zero mean.  For example, for the $x$ velocity component, we write
\begin{eqnarray*}
u(x,X,y,z,t,T)&=&\bar{u}(X,y,z,T)\,+\,u'(x,X,y,z,t,T),
\end{eqnarray*}
where the overbar denotes the fast $(x,t)$ average. The motivation for averaging only in $x$ rather than over the entire horizontal ($x,z$) plane is that nontrivial spanwise structure of the streamwise-averaged streamwise velocity, associated with the streamwise streaks, plays a crucial role in the process that sustains the streamwise rolls.  With this decomposition, the mean equations can be expressed as
\begin{eqnarray}
\epsilon\partial_T\bar{u}+\epsilon\partial_X\left[\bar{u}\bar{u}+\overline{u'u'}\right]
+\nabla_\bot\cdot\left[\bar{\mathbf{v}}_\bot\bar{u}+\overline{\mathbf{v}_\bot'u'}\right]
&=&-\epsilon\partial_X\bar{p}+\epsilon\nabla_\bot^2\bar{u}+\epsilon^3\partial_X^2\bar{u}\nonumber\\
&&+\,\epsilon\frac{\sqrt{2}\pi^2}{4}\sin\left(\frac{\pi y}{2}\right),\\
\epsilon\partial_T\bar{\mathbf{v}}_\bot+\epsilon\partial_X\left[\bar{u}\bar{\mathbf{v}}_\bot+
\overline{u'\mathbf{v}_\bot'}\right]+\nabla_\bot\cdot\left[\bar{\mathbf{v}}_\bot\bar{\mathbf{v}}_\bot
+\overline{\mathbf{v}_\bot'\mathbf{v}_\bot'}\right]&=&-\nabla_\bot\bar{p}
+\epsilon\nabla_\bot^2\bar{\mathbf{v}}_\bot+\epsilon^3\partial_X^2\bar{\mathbf{v}}_\bot,\nonumber\\
{}\\
\epsilon\partial_X\bar{u}+\nabla_\bot\cdot\bar{\mathbf{v}}_\bot&=&0.
\end{eqnarray}

We proceed by positing an appropriate asymptotic expansion for the various fields.  To this end, we are motivated in part by the scaling behavior identified in \cite{Wang07} and recalled in \S\ref{INTRO} for lower-branch ECS although we emphasize that the resulting asymptotically reduced system is not limited to lower-branch states. For large $Re$ the rolls comprising the streamwise-averaged flow in the perpendicular plane are weak, $\mathit{O}(1/Re)$, relative to the deviation of the streamwise-averaged streamwise flow from the base laminar profile (i.e., the streaks). A \emph{closed} and asymptotically consistent reduced model may be obtained by further positing that the (streamwise-varying) fluctuations are similarly weak relative to the mean streamwise flow, an assumption that is consistent with the scalings reported in table~\ref{scaling}.  Thus, we expand the velocity components and pressure as follows:
\begin{eqnarray}
\label{asump1}
u&\sim&\left(\bar{u}_0+u_0'\right)\,+\,\epsilon\left(\bar{u}_1+u_1'\right)+\ldots,\\
\mathbf{v}_\bot&\sim&\epsilon\left(\bar{\mathbf{v}}_{1\bot}+\mathbf{v}_{1\bot}'\right)\,+\,
   \epsilon^2\left(\bar{\mathbf{v}}_{2\bot}+\mathbf{v}_{2\bot}'\right)+\ldots,\\
\label{asump3}
p&\sim&\left(\bar{p}_0+p_0'\right)\,+\,\epsilon\left(\bar{p}_1+p_1'\right)\,+\,
   \epsilon^2\left(\bar{p}_2+p_2'\right)+\ldots
\end{eqnarray}
At $\mathit{O}(1)$, Eqs.~(\ref{VxEQN})--(\ref{CONTEQN}) imply
\begin{eqnarray}
\partial_t u_0' + (\bar{u}_0 + u_0') \partial_x u_0'&=&-\partial_x p_0',\\
0&=&-\nabla_\bot (\bar{p}_0 + p_0'),\\
\partial_x u_0'&=&0.
\end{eqnarray}
From this equation set we conclude that $u_0'\equiv 0$ and $p_0'\equiv 0$.  Note that $\bar{p}_0$, if nonzero, can only depend on $X$ and $T$; this term is set to zero for PCF and WF, but may be retained for flows driven by externally-imposed mean pressure gradients (such as plane Poiseuille flow). 

Using these leading-order results, Eq.~(\ref{VxEQN}) yields at $\mathit{O}(\epsilon)$
\begin{eqnarray}
\partial_t u_1' + \partial_T\bar{u}_0 + \bar{u}_0\partial_x u_1' + \bar{u}_0\partial_X\bar{u}_0 + \left[(\bar{\mathbf{v}}_{1\bot}+\mathbf{v}_{1\bot}')\cdot\nabla_\bot\right]\bar{u}_0 &=& - \partial_x p_1' - \partial_X \bar{p}_0 + \nabla_\bot^2\bar{u}_0\nonumber\\
&&+\,\frac{\sqrt{2}\pi^2}{4}\sin\left(\frac{\pi y}{2}\right).\label{Vx0EQN}
\end{eqnarray}
Averaging this equation over the fast $x$ and $t$ variables, and using the fact that $\bar{p}_0\equiv0$ for both PCF and WF, we obtain
\begin{eqnarray}
\partial_T\bar{u}_0 + \bar{u}_0\partial_X\bar{u}_0
+ \left(\bar{\mathbf{v}}_{1\bot}\cdot\nabla_\bot\right)\bar{u}_0
&=&-\partial_X \bar{p}_0 + \nabla_\bot^2\bar{u}_0 + \frac{\sqrt{2}\pi^2}{4}\sin\left(\frac{\pi y}{2}\right).\label{U0bar}
\end{eqnarray}
Subtracting Eq.~(\ref{U0bar}) from Eq.~(\ref{Vx0EQN}) yields an equation for the streamwise fluctuating velocity $u_1'$:
\begin{eqnarray}
\partial_t u_1' + \bar{u}_0\partial_x u_1' 
+ \left(\bar{\mathbf{v}}_{1\bot}'\cdot\nabla_\bot\right)\bar{u}_0
&=&-\partial_x p_1'.\label{U1prime}
\end{eqnarray}
At $\mathit{O}(\epsilon)$, the perpendicular momentum equation (\ref{VpEQN}) takes the form
\begin{eqnarray}
\partial_t \mathbf{v}_{1\bot}' + \bar{u}_0\partial_x\mathbf{v}_{1\bot}'
&=&-\nabla_\bot (\bar{p}_1 + p_1'),
\end{eqnarray}
from which we conclude that
\begin{eqnarray}
\nabla_\bot \bar{p}_1&=&0\label{P1bar}
\end{eqnarray}
and
\begin{eqnarray}
\partial_t \mathbf{v}_{1\bot}' + \bar{u}_0\partial_x\mathbf{v}_{1\bot}'
&=&-\nabla_\bot p_1'.\label{Vp1prime}
\end{eqnarray}
Finally, the $\mathit{O}(\epsilon)$ continuity equation requires
\begin{eqnarray}
\partial_X\bar{u}_0\,+\,\nabla_\bot\cdot\bar{\mathbf{v}}_{1\bot}&=&0\label{CONTbar}
\end{eqnarray}
and
\begin{eqnarray}
\partial_x u_1'\,+\,\nabla_\bot\cdot\mathbf{v}_{1\bot}'&=&0.\label{CONTprime}
\end{eqnarray}

To obtain a closed reduced system, we average the $\mathit{O}(\epsilon^2)$ perpendicular momentum equation,
\begin{eqnarray}
\partial_t\mathbf{v}_{2\bot}'&+&\partial_T (\bar{\mathbf{v}}_{1\bot} + \mathbf{v}_{1\bot}' )+\bar{u}_0
\partial_x\mathbf{v}_{2\bot}'+(\bar{u}_1 + u_1') \partial_x\mathbf{v}_{1\bot}'+\bar{u}_0
\partial_X (\bar{\mathbf{v}}_{1\bot} + \mathbf{v}_{1\bot}')\nonumber\\
&+&\left[(\bar{\mathbf{v}}_{1\bot} + \mathbf{v}_{1\bot}' ) \cdot\nabla_\bot\right] (\bar{\mathbf{v}}_{1\bot} + \mathbf{v}_{1\bot}')\,=\,
-\nabla_\bot (\bar{p}_2 + p_2')\nonumber\\
&+&\left(\partial_x^2+\nabla_\bot^2\right) (\bar{\mathbf{v}}_{1\bot} + \mathbf{v}_{1\bot}' ),
\end{eqnarray}
and obtain, after averaging and using Eq.~(\ref{CONTprime}), an equation for the evolution of $\bar{\mathbf{v}}_{1\bot}$:
\begin{eqnarray}
\partial_T\bar{\mathbf{v}}_{1\bot}+\partial_X\left[\bar{u}_0\bar{\mathbf{v}}_{1\bot}\right]
+\nabla_\bot\cdot\left[\bar{\mathbf{v}}_{1\bot}\bar{\mathbf{v}}_{1\bot}+
\overline{\mathbf{v}_{1\bot}'\mathbf{v}_{1\bot}'}\right]&=&-\nabla_\bot\bar{p}_2
+ \nabla_\bot^2\bar{\mathbf{v}}_{1\bot}.\label{Vp1bar}
\end{eqnarray}
The resulting set of equations forms a {\it closed} system of equations valid in the limit $\epsilon\rightarrow 0$. This equation set is to be solved subject to the mean and fluctuating boundary conditions obtained by applying a similar decomposition to the conditions (\ref{boundary}):
\begin{equation}
\partial_y \bar{u}_0 = \bar{v}_1 = \partial_y \bar{w}_1  = v_1' = 0;
\end{equation}
the boundary condition $\partial_y w_1'=0$ is omitted owing to the absence of viscous terms in the fluctuation equations.

\subsection{Structure and regularization of the reduced model}

For ease of reference, we collect here the key results of the asymptotic analysis.  Specifically, the multiscale reduced model consists of Eq.~(\ref{U0bar}), with $\bar{p}_0\equiv 0$, Eqs.~(\ref{Vp1bar}) and (\ref{CONTbar}),
\begin{eqnarray}
\partial_T\bar{u}_0 + \bar{u}_0\partial_X\bar{u}_0
+ \left(\bar{\mathbf{v}}_{1\bot}\cdot\nabla_\bot\right)\bar{u}_0
&=&\nabla_\bot^2\bar{u}_0 + \frac{\sqrt{2}\pi^2}{4}\sin\left(\frac{\pi y}{2}\right),\label{U0barEQN}\\
\partial_T\bar{\mathbf{v}}_{1\bot}+\partial_X\left[\bar{u}_0\bar{\mathbf{v}}_{1\bot}\right]
+\nabla_\bot\cdot\left[\bar{\mathbf{v}}_{1\bot}\bar{\mathbf{v}}_{1\bot}+
\overline{\mathbf{v}_{1\bot}'\mathbf{v}_{1\bot}'}\right]&=&-\nabla_\bot\bar{p}_2
+ \nabla_\bot^2\bar{\mathbf{v}}_{1\bot},\label{Vp1barEQN}\\
\partial_X\bar{u}_0\,+\,\nabla_\bot\cdot\bar{\mathbf{v}}_{1\bot}&=&0,\label{CONTbarEQN}
\end{eqnarray}
which govern the mean (i.e., fast $x$ and $t$ averaged) dynamics, and Eqs.~(\ref{U1prime}), (\ref{Vp1prime}) and (\ref{CONTprime}), 
\begin{eqnarray}
\partial_t u_1' + \bar{u}_0\partial_x u_1' 
+ \left(\bar{\mathbf{v}}_{1\bot}'\cdot\nabla_\bot\right)\bar{u}_0
&=&-\partial_x p_1'\,+\,\epsilon\nabla_\bot^2u_1',\label{U1primeEQN}\\
\partial_t \mathbf{v}_{1\bot}' + \bar{u}_0\partial_x\mathbf{v}_{1\bot}'
&=&-\nabla_\bot p_1'\,+\,\epsilon\nabla_\bot^2\mathbf{v}_{1\bot}',\label{Vp1primeEQN}\\
\partial_x u_1'\,+\,\nabla_\bot\cdot\mathbf{v}_{1\bot}'&=&0,\label{CONTprimeEQN}
\end{eqnarray}
which govern the fluctuating fields.  Physically, the averaged equations constrain the slow temporal and streamwise evolution of the streaks ($\bar{u}_0$) and rolls ($\bar{\mathbf{v}}_{1\bot}$). The presence of an effective Reynolds number equal to unity together with the absence of fast streamwise and temporal variation suggests that these equations should be more computationally tractable than the full Navier--Stokes equations at large $Re$. Indeed, if the slow streamwise ($X$) variation is suppressed, the averaged equations are spatially 2D and may be expected to exhibit quasi-laminar behavior.  Thus, deviations from the base laminar flow, if nonzero, are driven \emph{solely} by the fluctuation-induced Reynolds stress divergence in Eq.~(\ref{Vp1barEQN}); this correlation involves only the perpendicular fluctuating velocity field, all other Reynolds stress components being asymptotically smaller than the retained mean terms.  

Presuming fluctuation gradients remain $\mathit{O}(1)$, the fluctuating fields themselves evolve in accord with the equations governing the \emph{inviscid stability} of streamwise streaks (under the consistent approximation that the $\mathit{O}(1/Re)$ rolls may be neglected). In particular, spanwise inflections in the profile of $\bar{u}_0(X,y,z,T)$ may be expected to give rise to an $x$-varying 3D instability whose primary effect will be to re-energize the streamwise rolls through the Reynolds stress term, in accord with the self-sustaining process theory \citep{Waleffe97}.  As explicitly demonstrated in \cite{Wang07}, the fluctuation (or wave) fields, which are necessarily steady (neutral) for equilibrium ECS, exhibit a critical layer structure along the isosurface $\bar{u}_0(y,z)$=$0$. In the neighborhood of the critical layer, the fluctuation gradients are large, resulting in a distinct leading-order dominant balance of terms involving diffusion. Unlike \cite{Hall10}, we choose to avoid the intricacies associated with carrying out a systematic matched asymptotic analysis to address the critical layer singularity.  Rather, as discussed in \S\ref{INTRO}, we proceed simply, but effectively, by retaining the formally small diffusion terms in (\ref{U1primeEQN}) and (\ref{Vp1primeEQN}); the retention of these terms may be justified by appeal to the method of \emph{composite asymptotic approximations} or to the related method of \emph{composite asymptotic equations} \citep{Giannetti06}.  A similar regularization was employed in the recent work of \cite{Blackburn13}.

It is significant that the fluctuation equations (\ref{U1primeEQN})--(\ref{CONTprimeEQN}) do not mix $x$ modes, a fact we exploit in our computations of ECS for WF using the reduced system. Specifically and in accord with the scalings given in table~\ref{scaling}, we retain only the fundamental streamwise Fourier mode for each fluctuation field, and write
\begin{eqnarray}
u_1'(x,y,z,t)&=&\hat{u}_1(y,z,t)e^{i\alpha x}\,+\,c.c.,
\end{eqnarray}
where $\alpha\equiv 2\pi/L_x$ is the dimensionless fundamental streamwise wavenumber and $c.c.$ denotes the complex conjugate; similar expressions are written for $v_1'$, $w_1'$ and $p_1'$. In the following, we drop the hat over fluctuating variables for brevity of notation. In very long domains a nearly continuous band of modes with similar streamwise wavenumbers will be neutral or very weakly damped, leading to a description of the flow in terms of an evolving linear superposition of these modes exhibiting a slowly-varying envelope. This evolution will in turn drive slow streamwise modulations of the mean fields through the Reynolds stress divergence term in Eq.~(\ref{Vp1barEQN}). A mechanism of this type may provide an explanation for the streamwise localization of ECS observed in a variety of plane parallel shear flows \citep{Schneider10}, further attesting to the value of the reduced structure identified here.

In the first instance, however, slow streamwise variations can be suppressed.  The resulting averaged equations (\ref{U0barEQN})--(\ref{CONTbarEQN}) can then be further simplified by introducing a streamwise-invariant streamfunction $\phi_1$: $\bar{v}_1 = -\partial_z \phi_1$, $\bar{w}_1 = \partial_y \phi_1$, yielding the streamwise-invariant vorticity $\omega_1 = \nabla_{\perp}^2 \phi_1$. Consequently, the averaged system can be expressed as
\begin{eqnarray}
\label{rdc1}&\partial_T u_0 + J(\phi_1,u_0) = \nabla_{\perp}^{2} u_0 + \frac{\sqrt{2} \pi^2}{4} \sin(\pi y /2),\\
\label{rdc2}&\partial_T \omega_1 + J(\phi_1,\omega_1) + 2 (\partial_{yy}^2 - \partial_{zz}^2) \left( \mathcal{R}(v_1 w_1^*) \right) + 2 \partial_y \partial_z (w_1 w_1^* - v_1 v_1^*) = \nabla_{\perp}^2 \omega_1,
\end{eqnarray}
where $J(\phi_1,f) = \partial_y \phi_1 \partial_z f - \partial_z \phi_1 \partial_y f$ and $\mathcal{R}$ denotes the real part. In writing these equations, we have dropped the overbar on the mean streamwise velocity component $u_0$, again for notational brevity. The fluctuation equations can also be simplified, in particular by taking the divergence of Eqs.~(\ref{U1primeEQN}) and (\ref{Vp1primeEQN}) and using Eq.~(\ref{CONTprimeEQN}) to obtain a Helmholtz equation for the pressure $p_1$. The resulting fluctuation equations can be written in the form
\begin{eqnarray}
\label{rdc3}&(\alpha^2 - \nabla_{\perp}^2) p_1 = 2 i \alpha (v_1 \partial_y u_0 + w_1 \partial_z u_0),\\
\label{rdc4}&\partial_t \mathbf{v}_{1\bot} + i \alpha u_0 \mathbf{v}_{1\bot} = -\nabla_{\bot} p_1 + \epsilon \nabla_{\perp}^2 \mathbf{v}_{1\bot}.
\end{eqnarray}
The boundary conditions at $y=\pm 1$ are
\begin{equation}
\label{rdcbc}
\partial_y u_0 = \omega_1 = \partial_y \phi_1 =v_1 = \partial_y w_1 = 0,
\end{equation}
together with periodic boundary conditions in $z$. Observe that $u_1$ does not appear in these equations although it can be recovered from Eq.~(\ref{CONTprimeEQN}).

Equations (\ref{rdc1})--(\ref{rdc4}) capture the self-sustaining process explicitly: the rolls $\omega_1$ ($\phi_1$) deform the structure of the streamwise velocity $u_0$ to generate streaks as described by Eq.~(\ref{rdc1}). These streaks lead to the formation of a fluctuating structure $\mathbf{v}_{1\bot}$ through the advection term in Eq.~(\ref{rdc4}). Lastly, these fluctuations feed the rolls through the Reynolds stresses in Eq.~(\ref{rdc2}). The reduced model (\ref{rdc1})--(\ref{rdc4}) thus isolates the self-sustaining process described by \cite{Waleffe97}.  

Before describing in the next section a numerical algorithm for computing ECS admitted by the reduced system, we clarify the relation of our reduced model to the formulations of \cite{Hall10} and \cite{Blackburn13}.  The elegant analysis of \cite{Hall10} clearly demonstrates that, for the Nagata--Busse/Clever--Waleffe lower-branch equilibrium solution in PCF, the amplitude of the fluctuating fields actually scales as $Re^{-7/6}$ -- not $Re^{-1}$ -- away from the critical layer. Within the critical layer, the fluctuation velocity components tangent to the critical layer are amplified, becoming $\mathcal{O}(Re^{-5/6})$ -- again rather than $\mathcal{O}(Re^{-1})$ as prescribed here. Thus, in the limit $Re\to\infty$, the fluctuation-induced forcing of the rolls is asymptotically confined to the critical layer, justifying the jump condition formulation derived in \cite{Hall10}.  Given these non-integer scalings, several comments are in order regarding implications for the validity and utility of the present modeling approach.  First, it should be emphasized that our reduced equations retain all of the terms required to capture the limiting physics; however, for very large values of $Re$, our numerics must implicitly capture the weak $Re$ dependence of the fluctuations (which are a factor $Re^{1/6}$ larger within the critical layer but $Re^{-1/6}$ smaller outside). Indeed, the same issue arises in the hybrid method of \cite{Blackburn13}, in which the fluctuation fields are formally scaled as $Re^{-7/6}$ throughout the entire physical domain.  Secondly, for more moderate values of $Re$ that are still numerically much larger than unity (e.g., $Re=\mathcal{O}(10^3)$), $Re^{-1/3}$ is not particularly large, so the distinction between the fluctuation amplitude within and outside the critical layer is somewhat blurred.  But it is precisely this Reynolds number regime that is of interest in transition studies. Moreover, the results of \cite{Wang07} suggest that the scale separation between the fluctuation/roll fields and the streak field, on which the reduced model is founded, is already evident at these values of the Reynolds number. In addition, there may be other, perhaps non-equilibrium (e.g., a periodic-orbit) ECS that do \emph{not} exhibit critical layer structure but that can nevertheless be captured by our self-consistent asymptotically-reduced model.

\section{Numerical strategy}\label{NUMERICS}

A common approach for computing edge solutions is to apply a so-called edge-tracking algorithm, which only requires use of a time-stepper for the equations \citep{Skufca06}. In the present case, however, implementation of this technique is complicated by the occurrence of two time scales $t$ and $T=\epsilon t = t/Re$ in our reduced equations: the mean variables $u_0, \omega_1$ evolve on the long time scale $T$, while the fluctuating variables $\mathbf{v}_{1\bot}$ and $p_1$ vary on the fast time scale $t$. In the small $\epsilon$ limit, the slow mean variables are quasi-steady during the evolution of the fast fluctuating variables. Thus, the fluctuation equations (\ref{rdc3})--(\ref{rdc4}) are effectively quasi-linear, and the mean variables only respond to the evolution of the fluctuations on a longer time scale. Consequently, naive time integration of the system (\ref{rdc1})--(\ref{rdc4}) leads either to decay to the trivial solution or to unarrested exponential growth of the fluctuations, with the mean variables unable to provide feedback on the required time scale.  We note in passing that a multiscale time-stepping strategy, as provided by heterogeneous multiscale methods or gap-tooth and projective integration schemes \citep{Vandeneijnden07}, may be feasible since the fluctuation system is then integrated only for short `bursts'. Rather than pursuing such an approach to time integration, we summarize below a procedure that avoids edge tracking altogether by treating the fluctuation system as an eigenvalue problem \citep{BeaumeGFD12}; a similar strategy was followed by \cite{Hall10} and \cite{Blackburn13}.

In the following, we consider a two-dimensional domain $\mathcal{D}$ initially of size $L_y \times L_z = 2 \times \pi$, where $L_y=2$ is the (dimensionless) distance between the walls and $L_z=\pi$ is the (dimensionless) imposed period in the spanwise direction, and set the streamwise wavenumber $\alpha = 0.5$. For PCF this domain yields the least unstable lower branch solution \citep{Schneider08}. The two-dimensional domain is meshed using equidistributed points and the solutions expressed in a Fourier basis. The equations are treated in spectral space using the Fast Fourier Transform in the periodic direction $z$ and either the Fast Cosine Transform-I or Fast Sine Transform-I in the wall-bounded direction $y$, depending on the boundary conditions: $u_0$ and $w_1$ are expanded in a cosine basis while $\omega_0$ and $v_1$ are expanded in a sine basis \citep{FFTW05}. All spatial derivatives are computed pseudospectrally in physical space. The usual 2/3 dealiasing is applied for quadratic nonlinearities to avoid mode contamination by spectral convolution.

By analogy with PCF we seek solutions that are shift-reflect-symmetric, i.e., solutions that are invariant under the operation $[u,v,w](x,y,z) = [u,v,-w](x+L_x/2,y,-z)$, where $L_x$ is the imposed period in the streamwise direction -- in our case $L_x = 4 \pi$. Within the reduced model framework, this operation becomes $[u_0,\omega_1,v_1,w_1](y,z) = [u_0,-\omega_1,-v_1,w_1](y,-z)$. In addition, the solutions can be translated in $x$: $[u,v,w](x,y,z)$ $\rightarrow [u,v,w](x+\ell,y,z)$, where $\ell$ is an arbitrary real quantity. Within our approach this symmetry corresponds to $[u_0,\omega_1,v_1,w_1](y,z) \rightarrow [u_0,\omega_1,\mathcal{R}(v_1)\cos(\ell')-\mathcal{I}(v_1)\sin(\ell') + i (\mathcal{R}(v_1)\sin(\ell')+\mathcal{I}(v_1)\cos(\ell')),\mathcal{R}(w_1)\cos(\ell')-\mathcal{I}(w_1)\sin(\ell') + i (\mathcal{R}(w_1)\sin(\ell')+\mathcal{I}(w_1)$ $\cos(\ell'))](y,z)$, where $\mathcal{R}(\cdot)$ (resp. $\mathcal{I}(\cdot)$) denotes the real (resp. imaginary) part and $\ell'\equiv\alpha\ell$. 

In the next subsection we propose an iterative strategy for obtaining a good initial condition for the successful convergence of a Newton iteration used to obtain ECS and then present a preconditioner for the reduced model (\ref{rdc1})--(\ref{rdc4}) that allows efficient continuation of the lower branch. Upper branch solutions can also be continued, but in a more \emph{ad hoc} manner, as explained further below.

\subsection{The initial iterate}\label{initial}

We start by decoupling the mean variables $u_0$ and $\omega_1$ that evolve slowly from the fluctuations $v_1$ and $w_1$ that evolve more rapidly. We thus consider in succession the slow equations (\ref{rdc1}) and (\ref{rdc2}) in which the Reynolds stresses are fixed, and the fast equations (\ref{rdc3}) and (\ref{rdc4}) in which the quantity $u_0$ is maintained constant. Within this framework, the fluctuation equations are linear and autonomous. We take advantage of this structure by treating the fluctuation system as an eigenvalue problem, i.e., we seek solutions with exponential dependence in time: $\mathbf{v_{1\bot}}(y,z,t) \equiv \mathbf{{\hat v}_{1\bot}}(y,z) e ^{\lambda t}$ and $p_1(y,z,t) \equiv {\hat p}_1(y,z) e ^{\lambda t}$, where $\lambda$ is the growth rate of the fluctuations. This approach is critical as it provides more information on the fast dynamics than is available with time-steppers, which only determine the dominant modes. The basic idea is straightforward: if one of the fluctuating modes is marginal, it corresponds to a stationary solution of the fluctuation equations and if the associated mean variables are also stationary, then the combined mean/fluctuation fields comprise a stationary solution of the reduced system (\ref{rdc1})--(\ref{rdc4}).

The separate treatment of the mean and fluctuation problems implies that the scalar amplitude of the fluctuations in Eqs.~(\ref{rdc3}) and (\ref{rdc4}) is not fixed by the eigenvalue solve, but must be self-consistently determined as part of the iterative procedure.  We refer to this \emph{a priori} unknown scalar as $A$ and define it mathematically in \S\ref{RESULTS}.  Finding a solution of the problem (\ref{rdc1})--(\ref{rdc4}) is then equivalent to finding the correct fluctuation amplitude $A$ for which, given stationary mean variables $u_0$ and $\omega_1$, there exists a fluctuating mode with vanishing growth rate. To obtain a good first approximation to an ECS for subsequent refinement and continuation via Newton iteration, we use the following multi-step iterative algorithm:
\begin{itemize}
\item[1.]{Arbitrarily choose the fluctuation amplitude $A$}
\item[2.]{If the growth rate $\lambda$ of the fastest non-oscillatory growing (or slowest decaying) mode is nonzero:}
\begin{itemize}
\item[2.1.]{Compute the fastest non-oscillatory growing (or slowest decaying) fluctuating mode and its growth rate $\lambda$ from equations (\ref{rdc3}) and (\ref{rdc4})}
\item[2.2.]{Time-advance $u_0$ and $\omega_1$ to steady state using equations (\ref{rdc1}) and (\ref{rdc2})}
\item[2.3.]{Repeat steps 2.1 and 2.2 until a converged growth rate $\lambda(A)$ is obtained}
\end{itemize}
\item[3.]{Adjust $A$ to drive $\lambda(A)$ to zero by repeating steps 2.1--2.3.}
\end{itemize}

To use this algorithm, an initial condition for $u_0$ alone is required. Solutions of the eigenvalue problem in step 2.1 are obtained using the exponential power method and the package ARPACK \citep{arpack}. Note that this computation is equivalent to finding the stability of $u_0$ with respect to streamwise fluctuating perturbations of wavenumber $\alpha$. As we are interested in the least unstable, and hence the most dynamically influential, solutions, we focus on the fastest growing or slowest decaying mode. Other equilibria or periodic orbits may be found by looking at subsequent eigenvalues but this is outside the scope of the present paper. The time integration of Eqs.~(\ref{rdc1}) and (\ref{rdc2}) in step 2.2 is carried out using a semi-implicit third-order Runge--Kutta scheme \citep{Spalart91}. This step is very fast compared to the eigenvalue computation. To simplify the entire computation, we impose the shift-reflect symmetry during the eigenvalue search and the time integration of the mean equations.

\subsection{Preconditioned Newton method}\label{cont}

Following the computation of good approximates using the iterative algorithm just described, the final step is to converge these solutions to the desired accuracy. Typically, this is done using a Newton method for which an inner iteration is required to invert a certain Jacobian matrix.  In the present case, the use of such a method is not straightforward due to the poor conditioning of the Jacobian, and a suitable preconditioner is therefore required. This preconditioner is based on that originally proposed by Tuckerman \citep{Tuckerman89,Mamun95} and can be adapted to the reduced system (\ref{rdc1})--(\ref{rdc4}), as we now describe. We consider a generic system of the form
\begin{equation}
\label{generic}
\gamma_t \partial_t U = N(U) + \gamma_D L U,
\end{equation}
where $U$ is the dependent variable, $N$ is a nonlinear operator, $L$ is a linear (Laplace) operator, with $\gamma_t$, $\gamma_D$ real constants. We look for stationary solutions and so aim to solve 
\begin{equation}
\label{rhs}
0 = N(U) + \gamma_D L U.
\end{equation}

The preconditioner introduced by Tuckerman is constructed from an implicit Euler scheme with time-step $\triangle t$ applied to Eq.~(\ref{generic}):
\begin{equation}
\label{impleuler}
U(t+\triangle t)=\left(I - \frac{\triangle t \, \gamma_D}{\gamma_t} \, L \right)^{-1} \left(U(t) + \frac{\triangle t}{\gamma_t} \, N(U(t))\right),
\end{equation}
where $U(t)$ stands for the value of $U$ at time $t$ and $I$ represents the identity operator. We note that by substracting $U(t)$ from expression (\ref{impleuler}), we obtain
\begin{equation}
\label{rhs2}
U(t+\triangle t)-U(t)= \frac{\triangle t}{\gamma_t} \left(I - \frac{\triangle t \, \gamma_D}{\gamma_t} \, L \right)^{-1} \Bigg( N(U(t)) + \gamma_D L U(t)\Bigg),
\end{equation}
where the right hand side of Eq.~(\ref{rhs}) is recovered and preconditioned by $P = I - \triangle t \, \gamma_D / \gamma_t \, L$.

A Newton method can be obtained by writing for iterate $U^{[k]}$ at iteration $k$
\begin{equation}
\label{newton}
N(U^{[k]}) + \gamma_D L U^{[k]} = J(U^{[k]}) \, \delta U,
\end{equation}
solving for $\delta U$ and correcting $U^{[k+1]} = U^{[k]} - \delta U$. In writing Eq.~(\ref{newton}), we have introduced $J(U^{[k]}) = \delta N(U^{[k]}) + \gamma_D L$, the Jacobian of the right hand side operator in Eq.~(\ref{generic}) with $\delta N(U^{[k]})\equiv\delta N / \delta U (U^{[k]})$.
On multiplying both sides of Eq.~(\ref{newton}) by $\triangle t / \gamma_t \, P^{-1}$, one obtains
\begin{equation}
\label{preprecon}
\frac{\triangle t}{\gamma_t} P^{-1} \left( N(U^{[k]}) + \gamma_D L U^{[k]} \right) = \frac{\triangle t}{\gamma_t} P^{-1} J(U^{[k]}) \, \delta U.
\end{equation}
The left hand side of Eq.~(\ref{preprecon}) can be obtained directly using Eq.~(\ref{rhs2}) by computing one implicit Euler time-step of the full equation (\ref{generic}) and substracting the initial condition. Moreover, applying the same method to Eq.~(\ref{generic}) linearized around $U^{[k]}$ we obtain a linearized version of Eq.~(\ref{rhs2}) that can then be used to calculate the right hand side of Eq.~(\ref{preprecon}). In the small $\triangle t$ limit, $P \approx I$ and the equation is not preconditioned while for sufficiently large $\triangle t$, we get the so-called Stokes preconditioner $P \approx \triangle t (\gamma_D / \gamma_t) L$. This preconditioning method is easy to implement as it only requires a first order implicit Euler time-integration scheme and its use is natural within matrix-free methods, where the Jacobian is not explicitly constructed. The Stokes preconditioner has been widely used in problems that are dominated by diffusion like coupled convection \citep{Bergeon02,Mercader06, Beaume13jfm}. In contrast, most shear flow studies are carried out at large Reynolds numbers and involve weakly diffusive flows. The required computations are then often performed without preconditioning \citep{channelflow,Duguetprivconv}.

In the present case, none of these approaches was efficient and we extended the previous preconditioning method to develop a mixed preconditioner. We note that the Jacobian is influenced by two terms: the linearized nonlinear term and the diffusive operator: $J(U) = \delta N(U) + \gamma_D L$. In the case of weakly diffusive flows, $\gamma_D \ll 1$ and the spectrum of the Jacobian is dominated by modes resulting from $\delta N(U)$. This is what happens in large Reynolds number studies. On the other hand, in the aforementioned convection problems $\gamma_D = \mathcal{O}(1)$. These problems are poorly conditioned due to the prominence of eigenvalues generated by the diffusion operator. This difficulty is natural and can be understood by the following heuristic argument: on a given Fourier grid, the condition number of a periodic Laplace operator scales in proportion to $k^2$, where $k$ is the largest wavenumber allowed. Thus accuracy is reached at the expense of poorer conditioning. This difficulty is typically handled by the use of Stokes preconditioning. In our reduced model, the mean equations (\ref{rdc1}) and (\ref{rdc2}) are diffusion-dominated: $\gamma_t = \epsilon^{-1}$ and $\gamma_D = 1$. According to the above considerations, these equations require Stokes-type preconditioning and we have found that $\triangle t = \epsilon^{-1} = Re$ (hence $\triangle t \gamma_D / \gamma_t = 1$ such that $P = I - L$) provides good results. The fluctuation equation (\ref{rdc4}), with Eq.~(\ref{rdc3}) solved as a preliminary step, is weakly diffusive with $\gamma_t = 1$ and $\gamma_D = \epsilon$ but was not efficiently solved without preconditioning. To improve the efficiency, we observed that the contribution of the nonlinear and the diffusive terms to the Jacobian depends strongly on the gradients within the critical layer. For the lower branch solution, the width of this layer scales like $(\alpha Re)^{-1/3}$, yielding a diffusion operator $(1/Re) \nabla_{\perp}^2 = \mathcal{O}(\alpha^{2/3} Re^{-1/3}) = \mathcal{O}(\alpha^{2/3} \epsilon^{1/3})$. We incorporate this scaling by setting $\triangle t = \xi^2 \alpha^{2/3} \epsilon^{-2/3}$, such that $P = I - \xi^2 \alpha^{2/3} \epsilon^{1/3} L$, where $\xi = \mathcal{O}(1)$ is a tuning constant. Several values of $\xi$ were tested and we adopted $\xi = 0.5$ for the computations that follow, unless stated otherwise.

To compute the desired ECS using a Newton search we embed the biconjugate-gradient-squared routine from NSPCG \citep{nspcg} within the Newton algorithm and impose a shift-reflect symmetry on the solutions of Eqs.~(\ref{rdc1})--(\ref{rdc4}).  Extra care is necessary to eliminate errors arising from the $x$-invariance of the solutions. Within our Fourier decomposition, translations correspond to the eigenvector $[u_0,\omega_1,v_1,w_1](y,z) = [0,0,iv_1,iw_1](y,z)$. The evaluation of the left side of the system corresponding to Eq.~(\ref{preprecon}) is carried out without any constraint while the right side is projected onto the space orthogonal to the above eigenvector, thereby removing the singularity of the Jacobian arising from translation invariance of the solution in $x$. The above procedure is implemented at each step of the Newton search and forms part of the continuation algorithm used to continue the converged solutions in parameter space.


\section{Exact coherent states}\label{RESULTS}

In this section, we present the ECS we have computed using the reduced model (\ref{rdc1})--(\ref{rdc4}) with stress-free boundary conditions (\ref{rdcbc}) at $y = \pm 1$ and periodic boundary conditions in $z$.

\subsection{Initial search}

We set $Re=400$ (or equivalently $\epsilon=1/400$) and employ the iterative strategy introduced in \S\ref{NUMERICS} on a $32 \times 32$ mesh. The accuracy of the results presented here is confirmed by computations on a $64 \times 64$ grid. We recall that the use of the iterative algorithm introduced in \S{3.1} only requires an initial condition on $u_0$. We generate such an initial condition by advecting the structureless Waleffe flow by a steady sinusoidal roll, $\omega_1(y,z) = 20 \sin(\pi/2 ) \sin(2 z)$ (or equivalently $\phi_1(y,z)=-5 (\pi^2/16 + 1)^{-1} \sin(\pi/2 y) \sin(2 z)$), integrating Eq.~(\ref{rdc1}) with a fixed $\phi_1$ until a steady state is reached. The roll structure and amplitude have been chosen such that the resulting initial condition on $u_0$ resembles the ECS in PCF \citep{Wang07,Schneider08,Hall10}. The resulting initial profile for the iterative algorithm is shown in figure \ref{initial}.
\begin{figure}
\centerline{\includegraphics[]{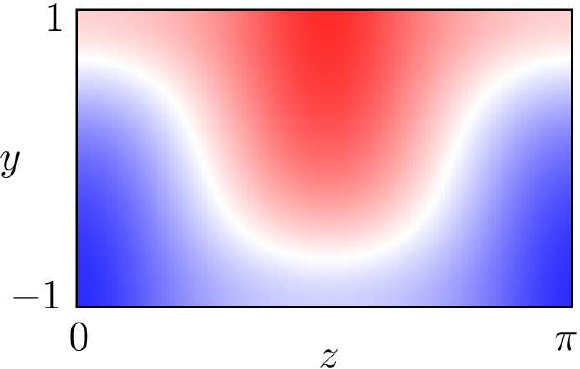}}
\caption{Initial condition for the iterative algorithm obtained by advecting/diffusing $u_0$ given a steady roll structure with $\omega_1(y,z) = 20\sin(\pi/2 y) \sin(2z)$. Red (blue) indicates positive (negative) values.}
\label{initial}
\end{figure}

\begin{figure}
\centerline{\includegraphics[width=10cm]{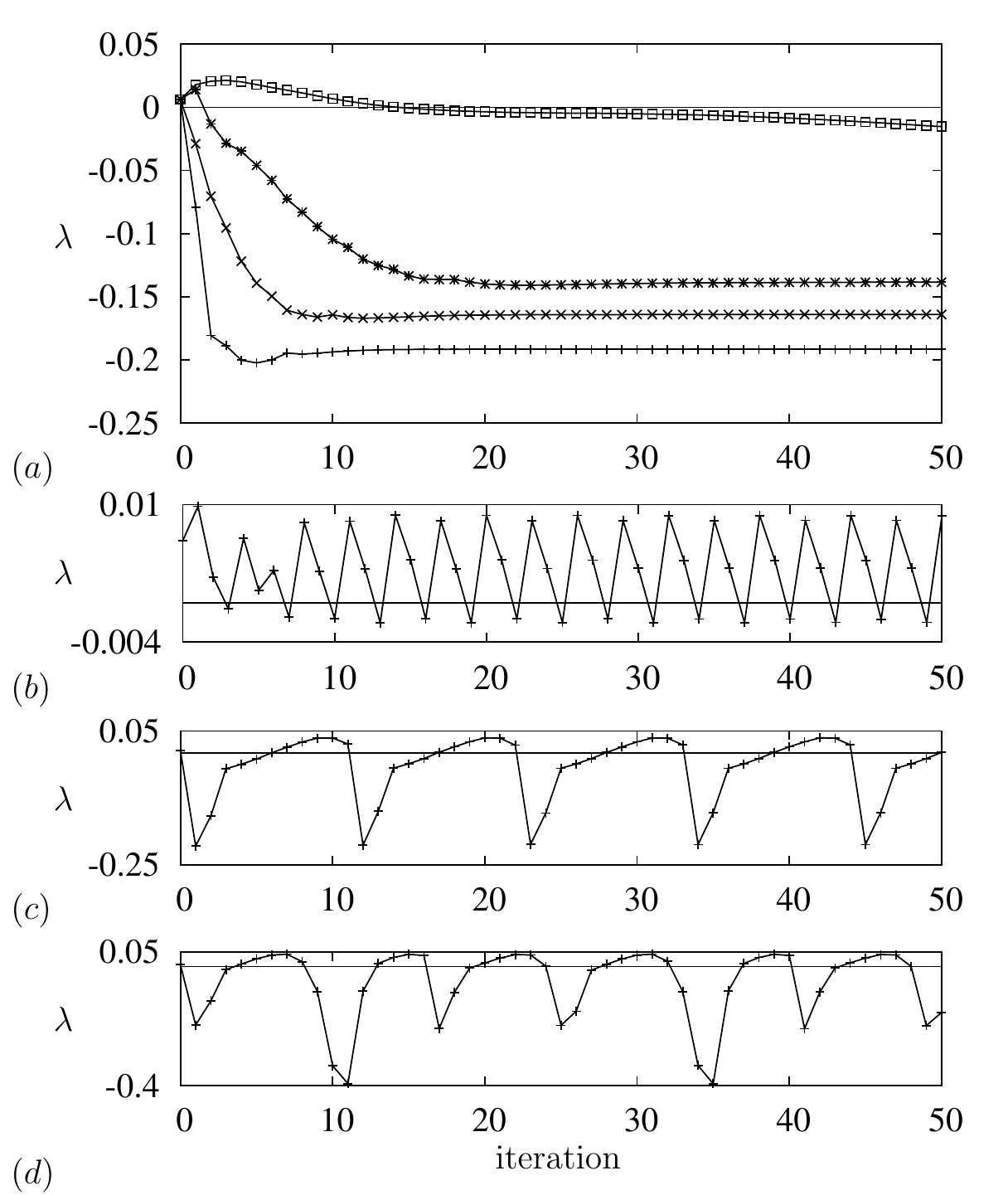}}
\caption{(a) Steady regime of the iterative algorithm ($A \le A_H \approx 6.81$). The curves represent the value of the largest real eigenvalue $\lambda$ plotted against the iteration number. Only the first $50$ iterations are shown and the curves correspond to $A=5$, $A=5.5$, $A=6$ and $A=6.5$ from bottom to top. The $A=6.5$ eigenvalue converges at a later iteration (not shown). The oscillatory regime ($A > A_H$) is represented in the same way for $A=6.9$ in (b), $A=7.5$ in (c) and $A=8$ in (d). In all cases, the algorithm is initialized using the initial condition shown in figure \ref{initial}.}
\label{regimes}
\end{figure}

We define the amplitude of the fluctuations numerically as the maximum value of any component of the in-plane fluctuating velocities on the meshgrid:
\begin{equation}
A=\max \left(|v_1(y_i,z_j)|;|w_1(y_i,z_j)| \right) \qquad \textrm{for $i=1,M$, $j=1,N$},
\end{equation}
where $y_i$ (resp. $z_j$) represents the $i$-th (resp. $j$-th) meshpoint in $y$ (resp. $z$), $M$ and $N$ are the number of points in $y$ and $z$, and $|f|=\sqrt{f_r^2+f_i^2}$ where the subscript $r$ (resp. $i$) denotes the real (resp. imaginary) part. We employed the iterative algorithm for different values of $A$ and observed two distinct regimes with different behavior of the leading real eigenvalue. In the first regime, observed for $A \le A_H \approx 6.81$, the leading real eigenvalue converges to $\lambda = \lambda_c (A)$. This regime is illustrated in figure \ref{regimes}(a).
For $A=0$ the solution converges to the Waleffe flow but as $A$ increases the converged flow departs from WF. For $A > A_H$, a second regime is present in which the algorithm does not converge but instead displays undamped oscillations (figures \ref{regimes}(b)--(d)). During the oscillations, the streaks $u_0$ alternately decay towards the trivial solution and then regrow into a more nonlinear structure that exaggerates the traits of the exact coherent state, a process that repeats in a periodic fashion. This process is straightforward for sufficiently small values of $A$ but becomes increasingly complex as $A$ is increased.  For example, the oscillations in $\lambda$ and the accompanying solution have a period of only $6$ iterations for $A=6.9$ (figure \ref{regimes}(b)) but $11$ iterations per oscillation for $A=7.5$ and $23$ iterations per oscillation for $A=8$ with $\lambda$ showing increasingly complex behavior (figures \ref{regimes}(c,d)).

The difference between these two regimes can be traced to the way in which $A$, the scalar amplitude of the fluctuations, enters Eq.~(\ref{rdc2}), where the Reynolds stress term has amplitude $A^2$. For a given fluctuation mode, increasing (decreasing) $A$ leads to a greater (lesser) forcing of the rolls $\omega_1$. The induced rolls deform the streaks $u_0$ generating a new eigenvalue problem for the fluctuations. Hence, $A$ is a forcing parameter that tunes one step in the self-sustaining process. If the forcing is too weak, the iterative algorithm relaxes to the trivial solution perturbed by a latent forcing induced by the non-vanishing fluctuations. If the forcing is too strong, the feedback from the mean variables (rolls $\omega_1$, then streaks $u_0$) is also too strong, causing overshooting of a potential ``steady" solution of the iterative algorithm. Since $A$ is fixed during the iteration process, successive overshoots occur, generating the observed oscillatory behavior. By analogy, one can think of a simple dynamical system which admits a stable steady solution at low $A$ before undergoing a supercritical Hopf bifurcation at $A = A_H$ to produce stable oscillations while the steady solution has become unstable. The variable in which the solution oscillates, the iteration number, is discrete which may account for the small departures from strictly periodic oscillations that can be observed in the data (figures \ref{regimes}(b)--(d)), as the period may vary continuously with $A$.

\begin{figure}
\centerline{\includegraphics[width=10cm]{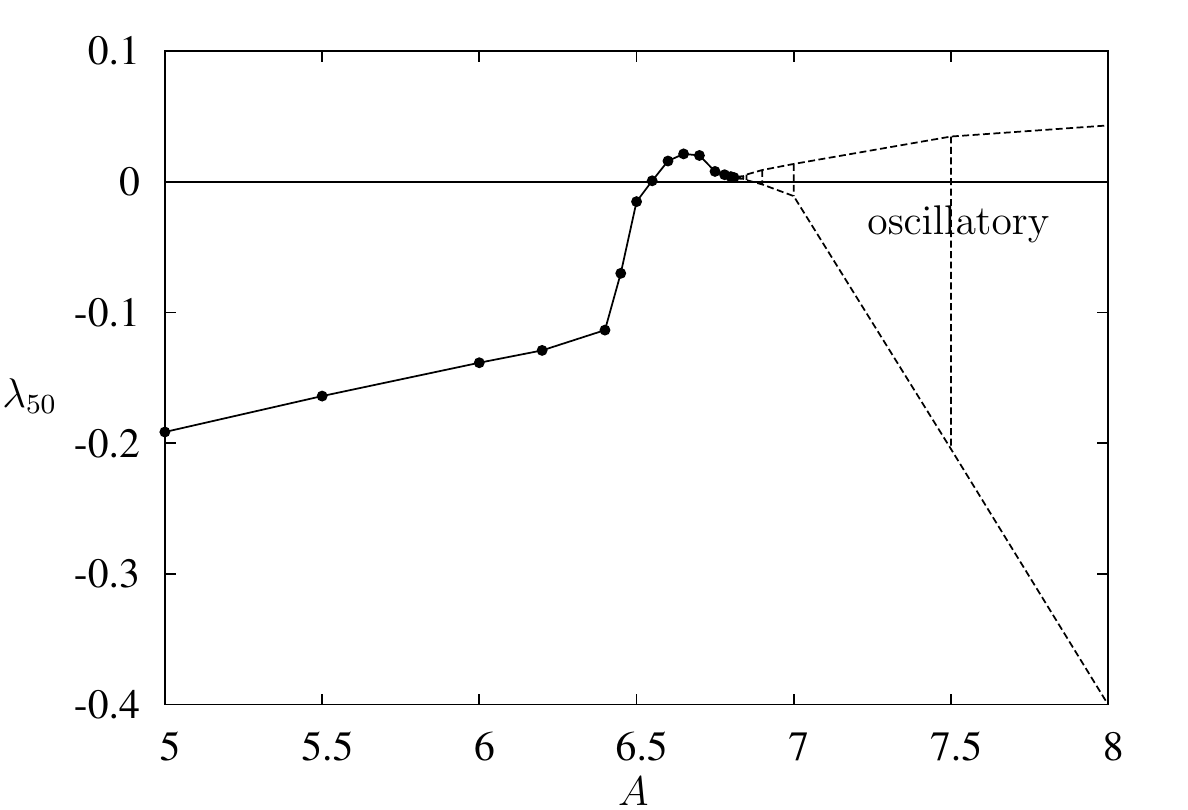}}
\caption{The leading stationary eigenvalue $\lambda_{50}$ after $50$ iterations of the algorithm in the steady regime ($A \le A_H \approx 6.81$) shown using a solid line with the dots indicating the value actually computed. Most of these eigenvalues decrease by less than $10^{-4}$ per iteration after $50$ iterations. The eigenvalues for $6.4 \le A \le 6.6$ (corresponding to the region of steepest increase in the figure) are not fully converged after $50$ iterations but do converge to slightly smaller values after a larger number of iterations. In fact convergence is not required as the aim is to generate a good initial condition for subsequent refinement by the preconditioned Newton method. The amplitude of the eigenvalue oscillations after $50$ iterations in the oscillatory regime ($A > A_H$) is indicated by vertical dashed lines at the values of $A$ actually used.}
\label{regime50}
\end{figure}

We look for ECS that are stationary, and so seek solutions with $\lambda_c$ close to zero. Figure \ref{regime50} shows the largest real eigenvalue after $50$ iterations $\lambda_{50}$ for values of $A$ spanning the interval $[5,8]$.
Although our results have been checked using a refined mesh ($64 \times 64$ modes) and the leading eigenvalue converged to $10^{-4}$ in most cases, we emphasize that convergence is not required at this stage as this algorithm is only intended to provide a good initial condition for a Newton iteration. The results reveal two possible candidates, corresponding to values of $\lambda_{50}$ close to $0$: $A_1 \approx 6.55$ and $A_2 = A_H \approx 6.81$. The state corresponding to $A_1$ is undoubtedly a good initial condition as $\lambda_{50}<0$ for $A < A_1$ and $\lambda_{50}>0$ for $A_2> A > A_1$. That corresponding to $A_2$ is seemingly less secure: for $A=6.81$, the converged eigenvalue is $\lambda_c(A=6.81) \approx 0.003191$ and approaches $0$ from above as $A$ is increased but the oscillatory regime is reached by $A=6.82$ leading to small amplitude oscillations close to but not crossing zero.

\subsection{Continuation in Reynolds number}

The regularized equations (\ref{rdc1})--(\ref{rdc4}) contain the parameter $\epsilon\equiv 1/Re$. The presence of this parameter allows us to continue the solutions in the Reynolds number even though the equations are formally valid only at large $Re$. While the solutions at finite $Re$ cannot be exact the self-consistency of the equations guarantees their usefulness. Moreover, as discussed further in \S\ref{Discussion}, the parameter $\epsilon$ can also be thought of as a homotopy parameter that allows us to locate other solutions to Eqs.~(\ref{rdc1})--(\ref{rdc4}) for large $Re$.

\begin{figure}
\centerline{\includegraphics[]{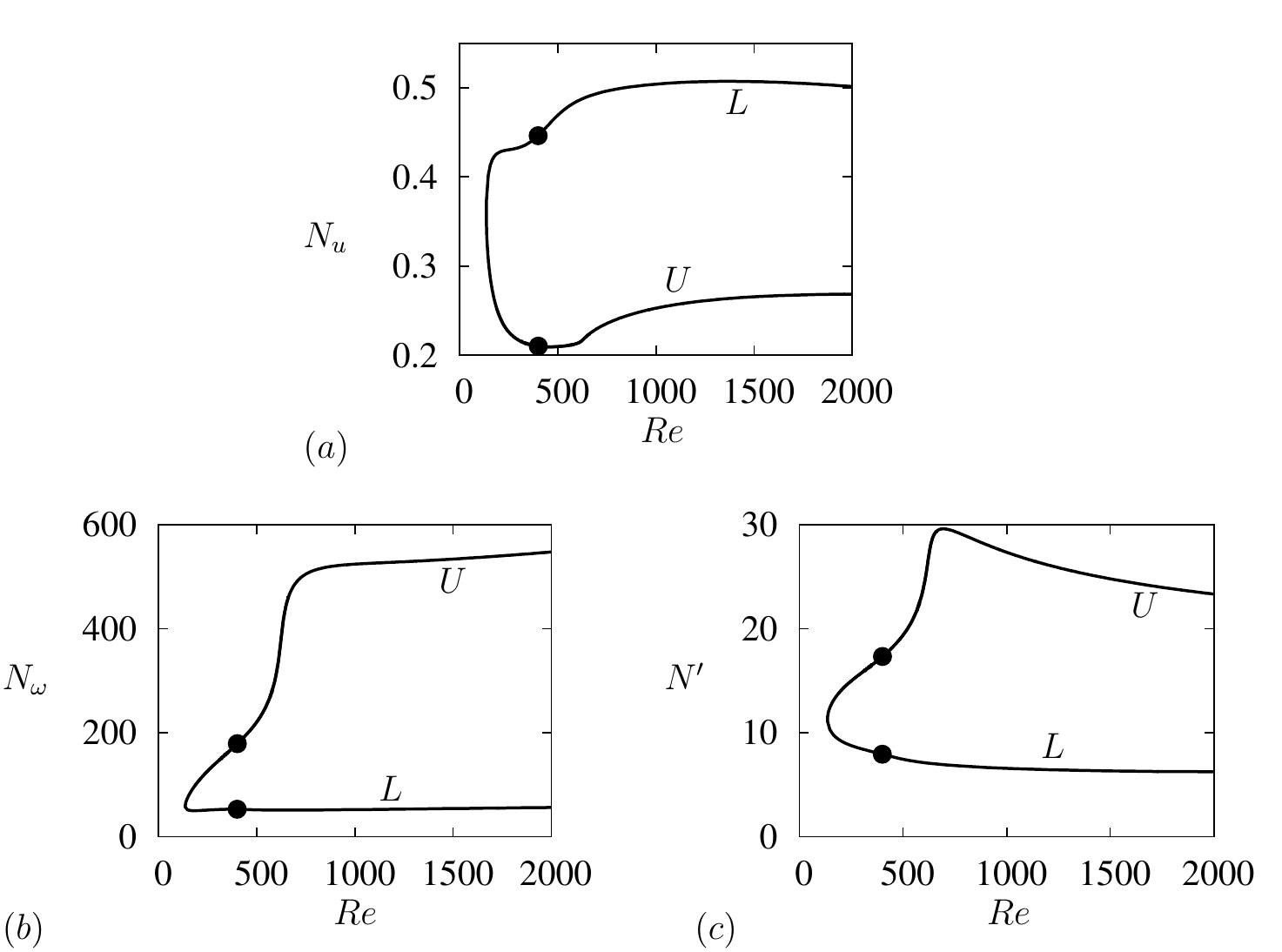}}
\caption{Bifurcation diagrams showing branches of exact coherent states as a function of the Reynolds number $Re$ obtained by continuation of converged solutions (solid dots) starting from approximates generated by the iterative algorithm with initial amplitude $A_1$ (converged to a lower branch state, indicated by $L$) and $A_2$ (converged to an upper branch state, indicated by $U$). (a) $N_u$, (b) $N_{\omega}$, (c) $N'$.}
\label{bifdiag}
\end{figure}

\begin{figure}
\centerline{\includegraphics[]{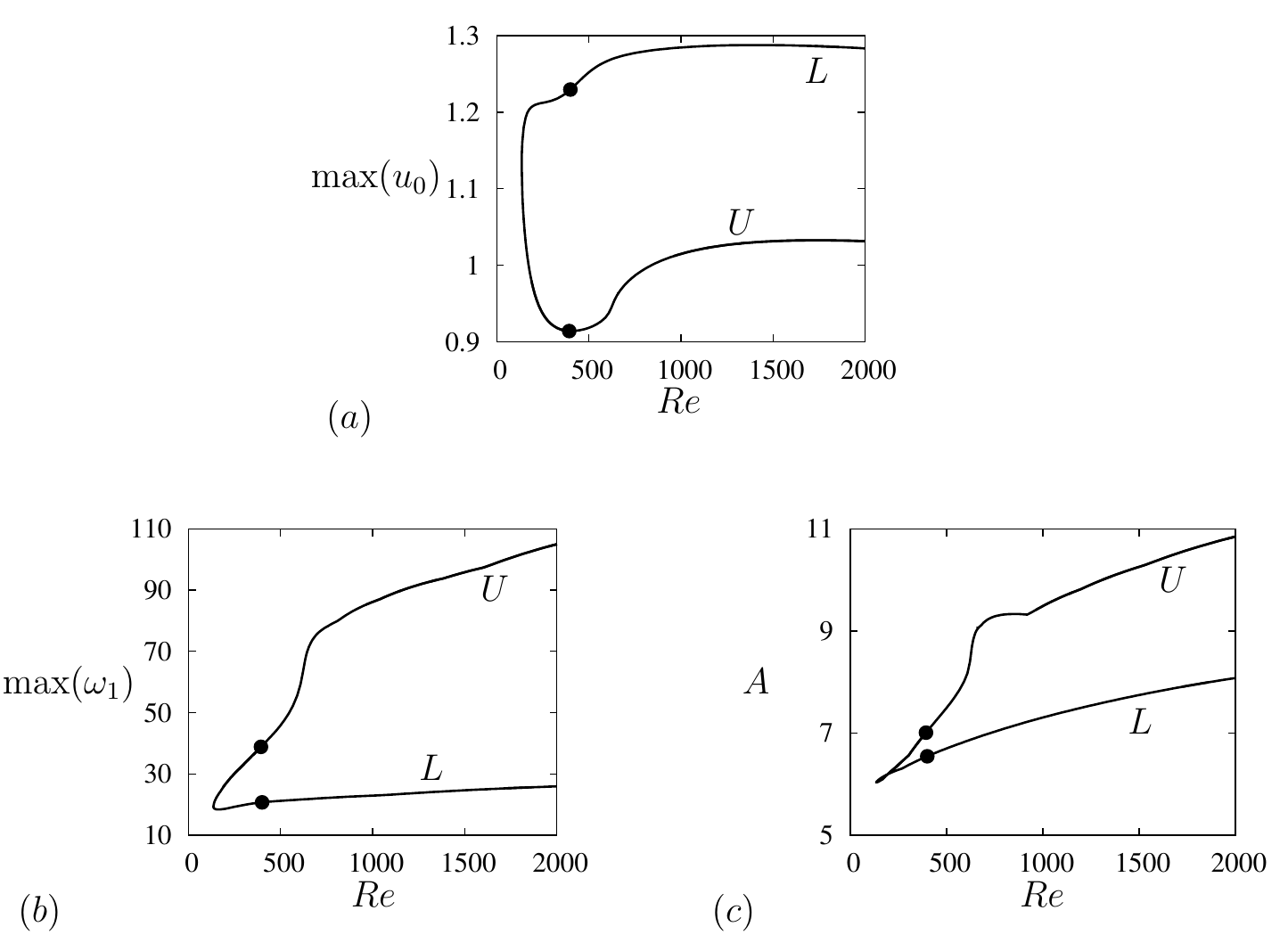}}
\caption{The same bifurcation diagrams as in figure \ref{bifdiag} rendered in terms of the maximum values of (a) the streamwise velocity $u_0$, (b) the vorticity $\omega_1$ and (c) the amplitude $A$ of the self-consistent fluctuation field.}
\label{flucampl}
\end{figure}

For this purpose we initialized a Newton search using the approximate solution with amplitude $A_1$ found in \S{3.1} and converged it to a lower branch state \citep{Beaumepreprint}. This solution was then continued to both larger and smaller Reynolds numbers using a $32 \times 64$ meshgrid; the results are displayed in figures \ref{bifdiag} and \ref{flucampl}.
The solution branches are plotted in six different ways. In figure \ref{bifdiag}(a) we present $N_u\equiv 2 E_u$, where $E_u\equiv\frac{1}{2D}\int_{\mathcal{D}} u_0^2(y,z) \, dy \, dz$ is the streamwise-invariant streamwise kinetic energy per unit volume and $D\equiv \int_{\mathcal{D}} dy \, dz$. In figure \ref{bifdiag}(b) we present $N_{\omega}\equiv 2\, Re^2 \, E_{\omega} \equiv \frac{1}{D}\int_{\mathcal{D}} \omega_1^2(y,z) \, dy \, dz $, a quantity related to the streamwise-invariant in-plane enstrophy per unit volume. In figure \ref{bifdiag}(c) we present the quantity $N' \equiv 2 Re^2 E'$, where $E' \equiv \frac{1}{2D \, Re^2}\int_{\mathcal{D}} (v_1^2+w_1^2) \, dy \, dz$ measures the streamwise-fluctuating in-plane kinetic energy per unit volume. The lower branch (labeled $L$) passes a saddle-node at $Re \approx 136$, giving rise to an upper branch (labeled $U$).  Figure \ref{flucampl} shows a projection of our solutions onto pointwise maxima of the corresponding quantities, thereby providing a complementary representation of the results. 

The emergence of the upper branch is rather unexpected since the reduced system (\ref{rdc1})--(\ref{rdc4}) was developed by appealing to lower branch scalings. Evidently the asymptotic procedure is sufficiently robust to capture both lower {\it and} upper solution branches. However, the computation of the upper branch is more delicate. As observed by \cite{Beaumepreprint}, upper branch solutions and their critical layer have a different spatial structure which dramatically increases the computational cost. To continue these solutions we used a $64 \times 128$ meshgrid and adjusted the preconditioner as necessary. Specifically, we started by testing a few values of $\triangle t$ (see \S{3.2}) and selected the most efficient one for continuation near the saddle-node. We then continued the upper branch until the algorithm failed. Each time this occurred we tested a few values of $\triangle t$ to determine the most suitable one, repeating this process as many times as necessary to continue the branch up to the desired value of the Reynolds number. Upper branch solutions can also be computed directly by starting from the approximate solution with amplitude $A_2$, obtained using the iterative scheme of \S{3.1}, and applying the Newton algorithm with the modified preconditioner described above. In particular, the upper branch is identified with the value of $A\approx7.04$ at which the (unstable) fixed point of the iterative process corresponds to zero eigenvalue. 


Some care is required in interpreting the values of the energies in figure \ref{bifdiag}, as $N_{\omega}$ and $N'$ are proportional to $Re^2$. The trivial solution, for which $N_u = 1$, $N_{\omega} = N' = 0$, represents the state of maximal transport and hence has the greatest kinetic energy. All other ECS are found to have a lower kinetic energy, as indicated by lower values of $N_u$. The quantities $N_{\omega}$ and $N'$, as well as the maximum of $\omega_1$ and the amplitude $A$, remain $\mathcal{O}(1)$ along the lower branch, reflecting the relevance of the asymptotic model. Interestingly, a similar observation can be made for the upper branch solution for which the fluctuations also remain $\mathcal{O}(1)$. However, possible departures from the assumed scaling may be observed in the enstrophy-related norm of the upper branch states. Indeed $N_{\omega} = \mathcal{O}(100)$, $\max(\omega_1) \sim 80-90$ at $Re = \mathcal{O}(1000)$, suggesting that the vorticity becomes larger and larger in an increasingly narrow region. However, these values, despite being relatively large at low $Re$ do not appear to increase sufficiently with $Re$ to violate the assumed form of the asymptotic expansion.

\begin{figure}
\centerline{\includegraphics[]{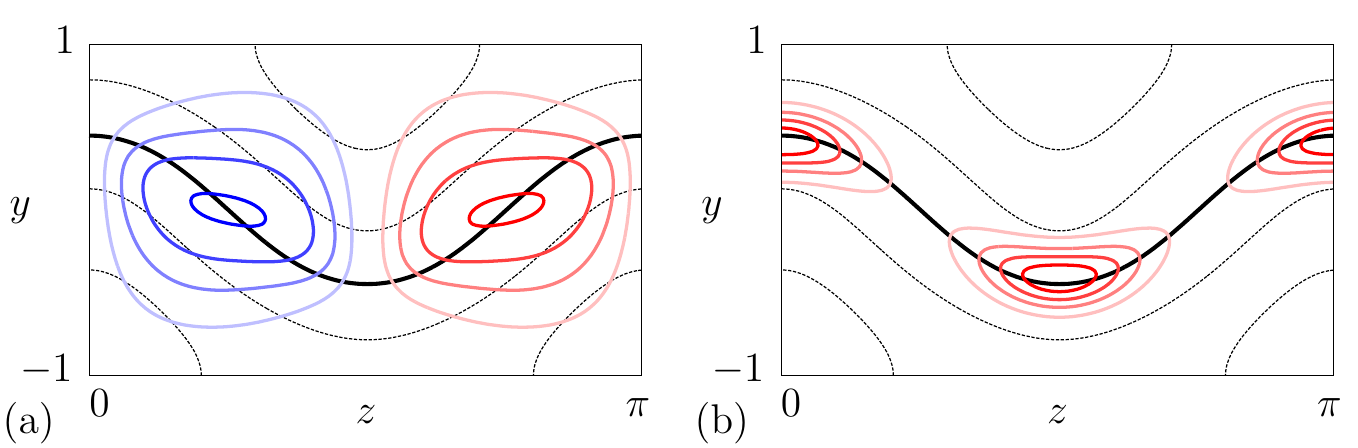}}
\caption{Lower branch solution computed at $Re \approx 1500$ represented by (a) contours of the streamwise-constant streamfunction $\phi_1$ and (b) contours of $|(v_1,w_1)|_{L_2}$ representing the amplitude of the in-plane fluctuations. In each of these plots, positive values are indicated in red while negative values are in blue; contours are equidistributed to give a sense of local gradients. Each contour plot is overlaid on the streak profile shown in black, with the solid line representing the critical layer $u_0 = 0$. Three-dimensional visualizations of the fluctuating variables are displayed in figure \ref{3dlower}.}
\label{lower}
\end{figure}

\begin{figure}
\centerline{\includegraphics[]{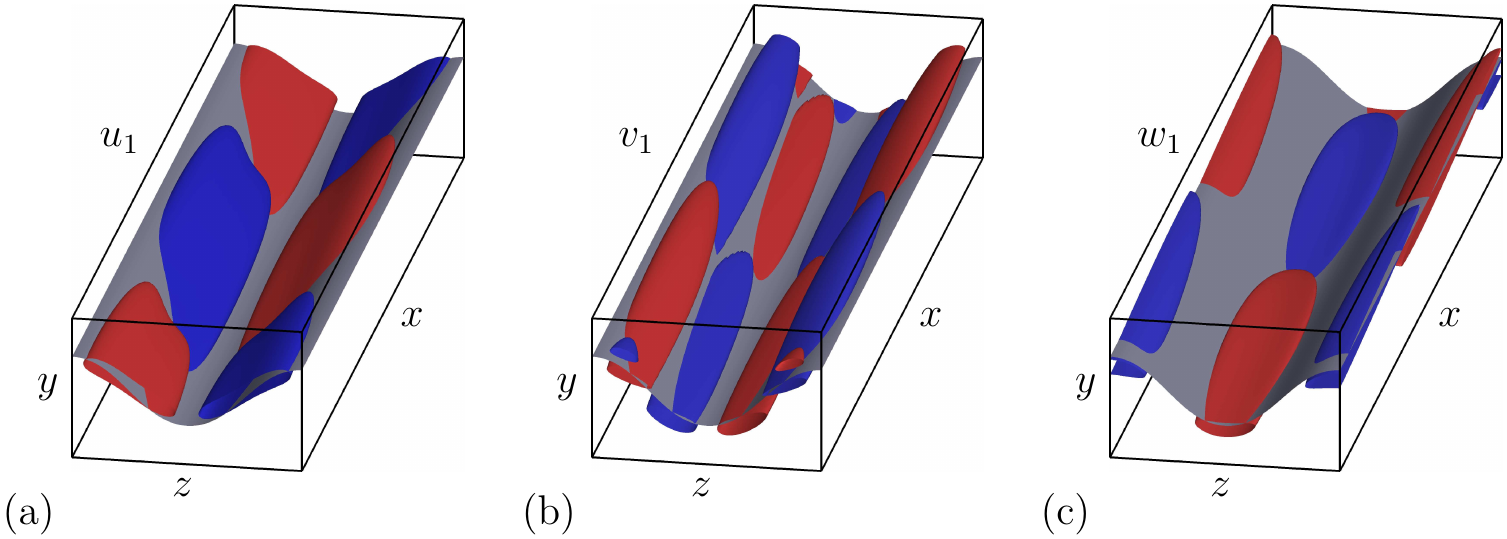}}
\caption{Three-dimensional rendition of the fluctuating flow associated with the lower branch solution at $Re \approx 1500$. The surfaces represented in color correspond to two equal and opposite values at half the maximum value of (a) the streamwise-fluctuating streamwise velocity $u_1$ , (b) the streamwise fluctuating wall-normal velocity $v_1$, and (c) the streamwise-fluctuating spanwise velocity $w_1$. Red color corresponds to positive values while blue corresponds to negative values. The grey surface shows the critical layer $u_0 = 0$.}
\label{3dlower}
\end{figure}

\begin{figure}
\centerline{\includegraphics[]{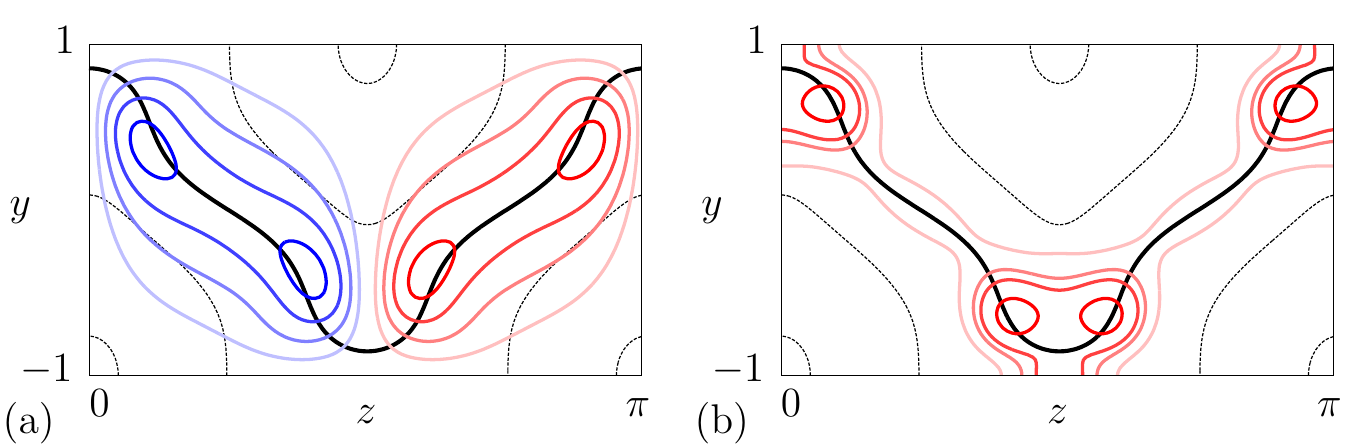}}
\caption{Same representation as in figure \ref{lower} but for the upper branch solution at $Re \approx 1500$. Three-dimensional visualizations of the fluctuating variables for this solution are displayed in figure \ref{3dupper}.}
\label{upper}
\end{figure}

\begin{figure}
\centerline{\includegraphics[]{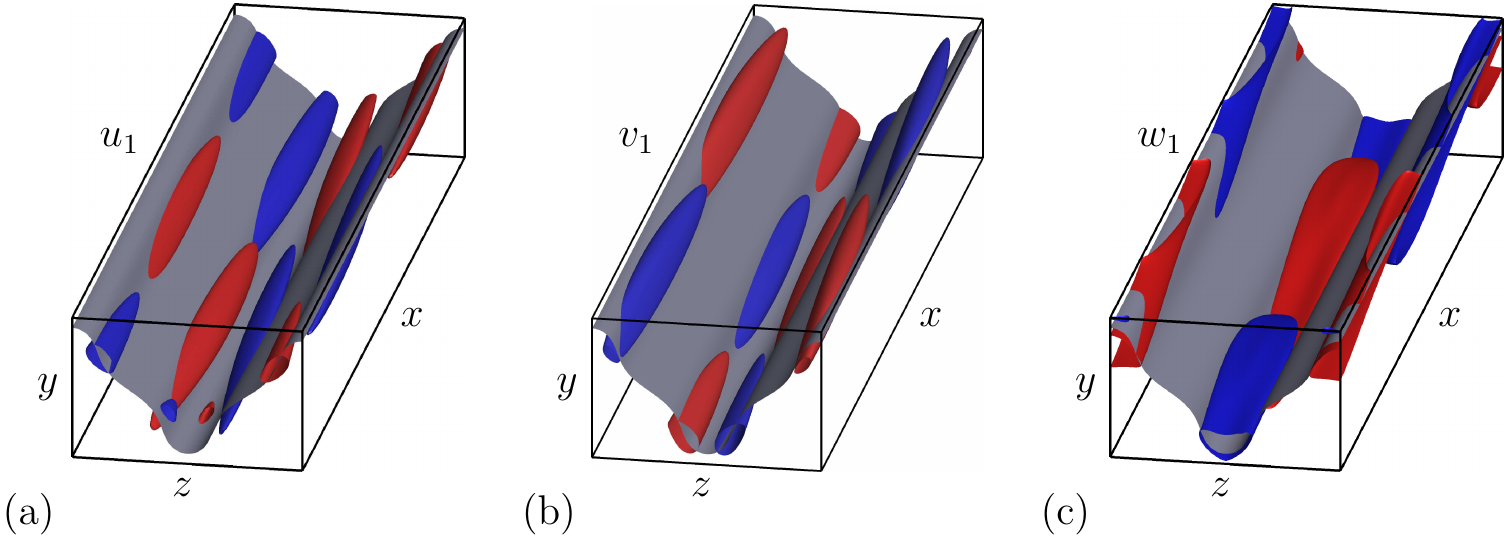}}
\caption{Same representation as in figure \ref{3dlower} but for the upper branch solution at $Re \approx 1500$. Intersections of non-zero fluctuations with the upper and lower walls can be observed in (c) and are allowed by the stress-free boundary conditions.}
\label{3dupper}
\end{figure}

Figure \ref{lower} depicts the lower branch solution at $Re \approx 1500$ using streamwise-averaged quantities while figure \ref{3dlower} provides a three-dimensional rendition of this solution. Figures \ref{upper} and \ref{3dupper} provide analogous representations of the upper branch solution at the same Reynolds number.
The lower branch solution possesses a smoothly undulating critical layer that is maintained by two nearly circular rolls (cf figure \ref{lower}(a)). This structure is supported by fluctuations accumulating in the critical layer. Figure \ref{lower}(b) shows that these fluctuations have a rapid variation in the direction perpendicular to the critical layer (its thickness being proportional to $(\alpha Re)^{-1/3}$) while slow variations are observed {\it along} the critical layer. The three-dimensional representations in figure \ref{3dlower} confirm these observations and shed some additional light on the streamwise dynamics of the lower branch solution. The streamwise-fluctuating streamwise velocity $u_1$ is essentially concentrated in the regions of stronger streamwise-invariant streamfunction $\phi_1$ (compare figure \ref{lower}(a) with figure \ref{3dlower}(a)) and therefore away from the crests of the critical layer. As a consequence of the incompressibility of the fluctuations (Eq.~\ref{CONTprimeEQN}), the in-plane fluctuating dynamics accumulate at the extrema of the critical layer, away from the location of the streamwise rolls, as documented in figure \ref{lower}(b). Figure \ref{3dlower} shows that at $x=0$ (defined arbitrarily as the front section in the figure), in the region around the lowest point of the critical layer, $z=\pi/2$, the fluid flows from left to right along the $u_0=0$ surface. The reverse occurs half a period downstream and at the highest point of the critical layer (located at the boundary of the (periodic) domain when $x=0$). 

In comparison to the lower branch solution, the upper branch solution has stronger variations along the critical layer, the extrema of which approach the top and bottom walls. This change in shape is a signature of stronger rolls. The resulting structure is shown in figure \ref{upper}(a), where by comparison with figure \ref{lower}(a), it is evident that the rolls are stretched diagonally and split, displaying a bimodal structure. This last feature is responsible for the sharper crests of the $u_0=0$ surface relative to that for the lower branch solution. This change in structure is reminiscent of the differences between lower and upper branch states in PCF (see figure 7 from \cite{Jimenez05}) and hints at the usefulness of our reduced model for states beyond the lower branch states for which it was developed. Associated with the bimodal structure in $\phi_1$ is a similar bimodal structure of the fluctuations which are now strongly localized on either side of the critical layer turning points. Figure \ref{upper} shows that as a result the location of the streamwise rolls almost coincides with the maxima of the fluctuation field, suggesting that the increased shear in the streamwise rolls suppresses fluctuations, with the location of the self-sustaining process moving towards the critical layer turning points. This evolution in turn implies that for the upper branch states the width of the critical layer depends strongly on location along the critical layer: the layer appears broader near its maximum deflection from $y=0$ and is substantially thinner in the intervals inbetween. In addition, the amplitude of both the rolls and the spanwise fluctuations peaks strongly in the vicinity of these turning points. Although we have not pursued this phenomenology further, the results suggest that in the limit $Re\to\infty$ the common assumption of uniform critical layer thickness may require reexamination, with the critical layer ``breaking up'' into something more akin to critical ``spots'', where most of the critical layer forcing is concentrated (figure \ref{upper}(b)).

\begin{figure}
\centerline{\includegraphics[]{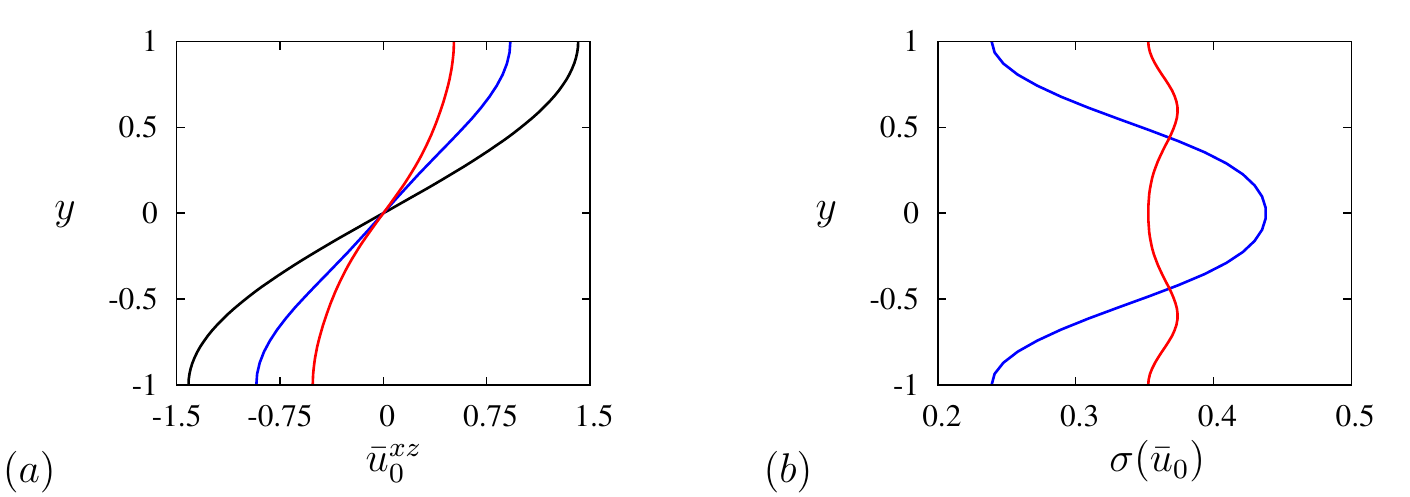}}
\caption{Streamwise velocity profiles at $Re \approx 1500$. (a) $(x,z)$-averaged streamwise velocity $\bar{u}_0^{xz}$ as a function of the wall-normal coordinate $y$. (b) Standard deviation $\sigma(\bar{u}_0)$ of the streamwise velocity $u_0$ from $\bar{u}_0^{xz}(y)$. The trivial solution is shown in black, the lower branch in blue and the upper branch in red. }
\label{profiles}
\end{figure}

These properties of the lower and upper branch solutions are reflected in the associated mean streamwise velocity profiles shown in figure \ref{profiles}. As expected, the ECS in each case reduces the shear across the layer. The reduction is less for the weaker lower branch ECS than for the upper branch ECS (figure \ref{profiles}(a)). Of particular interest is the standard deviation of the velocity from these profiles, $\sigma(\bar{u}_0,y) = \sqrt{\int_{z} (u_0(y,z) - \bar{u}_0^{xz})^2 dz}$, shown in figure \ref{profiles}(b). For the lower branch ECS $\sigma$ peaks at mid-height where the streamwise rolls are strongest and falls off quite strongly towards the walls at $y=\pm 1$ (figure \ref{profiles}(b)). In contrast, on the upper branch $\sigma$ is quite uniform across the layer, with a local minimum at mid-height, an effect that can be directly attributed to the stretching of the streamwise rolls along the critical layer and the location of their peak amplitude near regions of maximum deviation of the critical layer from $y=0$.

\subsection{Spectra}

\begin{figure}
\centerline{\includegraphics[]{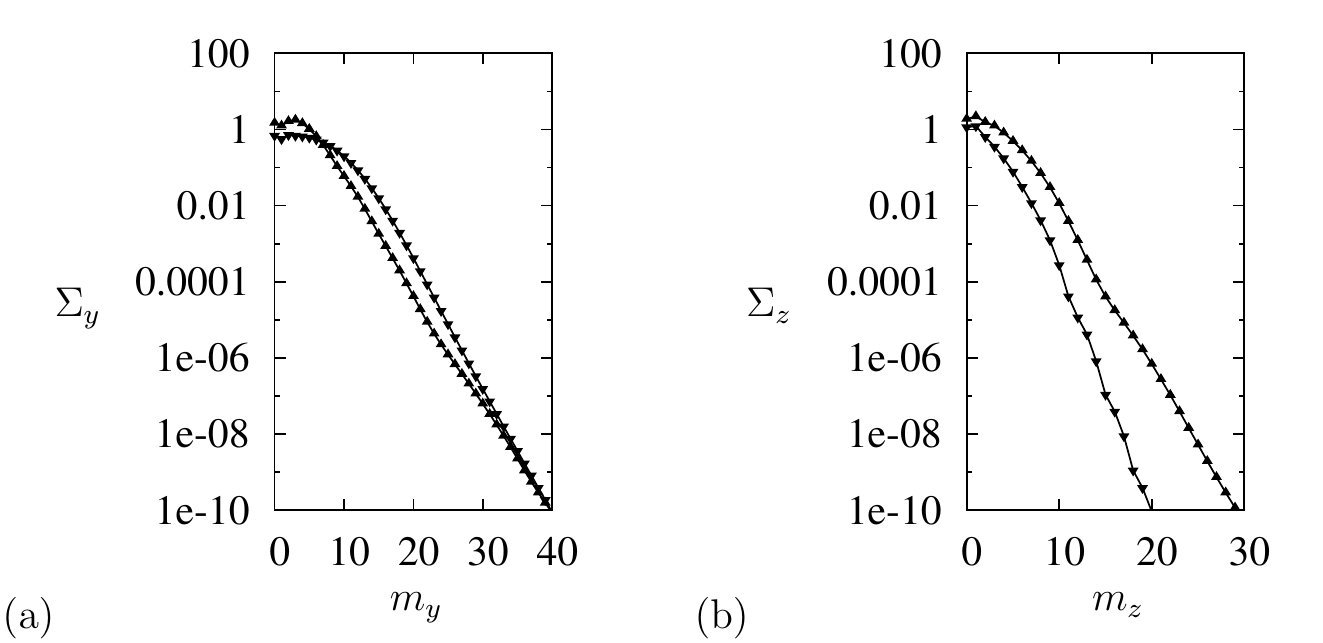}}
\caption{Spectra of the lower (downward triangles) and upper (upward triangles) branch solutions represented through the normalized partial sums $\sum_y (m_y)$ and $\sum_z (m_z)$ defined in Eqs.~(4.2) and (4.3).}
\label{fourier-spec}
\end{figure}

To check the accuracy of the solutions, we plot in figure~\ref{fourier-spec} the one-dimensional spectra of the fluctuation velocity in the wall-normal and spanwise directions. These are defined in terms of the normalized partial sums
\begin{eqnarray}
&\Sigma_y (m_y) = \frac{1}{2(M-1)N} \left(S^2(m_y,0) + \sum_{m_z=1}^{N} ( S^2(m_y,m_z) + S^2(m_y,-m_z))\right)^{1/2},\\
&\Sigma_z (m_z) = \frac{1}{2(M-1)N} \left(\sum_{m_y=0}^{M} S^2(m_y,m_z) \right)^{1/2},
\end{eqnarray}
where $S^2(m_y,m_z) = |v_1 (m_y,m_z)|^2 + |w_1 (m_y,m_z)|^2$ and $M$ (resp. $N$) is the maximum wavenumber in the $y$ (resp. $z$) direction. The quantity $\Sigma_y (m_y)$ (resp. $\Sigma_z (m_z)$) has been defined in such a way thay it is proportional to the sum of the amplitudes of the fluctuations with wavenumber $m_y$ (resp. $m_z$) in the $y$ (resp. $z$) direction. The plots confirm that the amplitude of the upper branch fluctuations is larger than that along the lower branch solutions. In addition, the spectra in the wall-normal direction decay exponentially at the same rate for both lower and upper branch states while the spectrum of the upper branch solution decays more slowly in the spanwise direction than that of the lower branch solution. These results reflect the fact that the scales of the wall-normal variation remains comparable as one goes from the lower branch to the upper one while the smallest spanwise scale shrinks. These results highlight the fact that the change in structure between the lower and upper branches is primarily associated with differences in the spanwise variation of the fields and inform the numerical requirements to compute these solutions accurately: while the wall-normal mesh can be designed independently of the solution sought with $30$ wavelengths sufficient at $Re = 1500$ (i.e., approximately $30$ modes in the cosine/sine basis, or $60$ modes in the complex Fourier basis), the number of points in the spanwise direction needs to be increased by a factor of about $1.5$ for upper branch states at $Re=1500$. Obviously, increasing the Reynolds number or any other factor that sharpens the critical layer impacts these requirements.

\subsection{ECS dependence on the domain size}

\begin{figure}
\centerline{\includegraphics[]{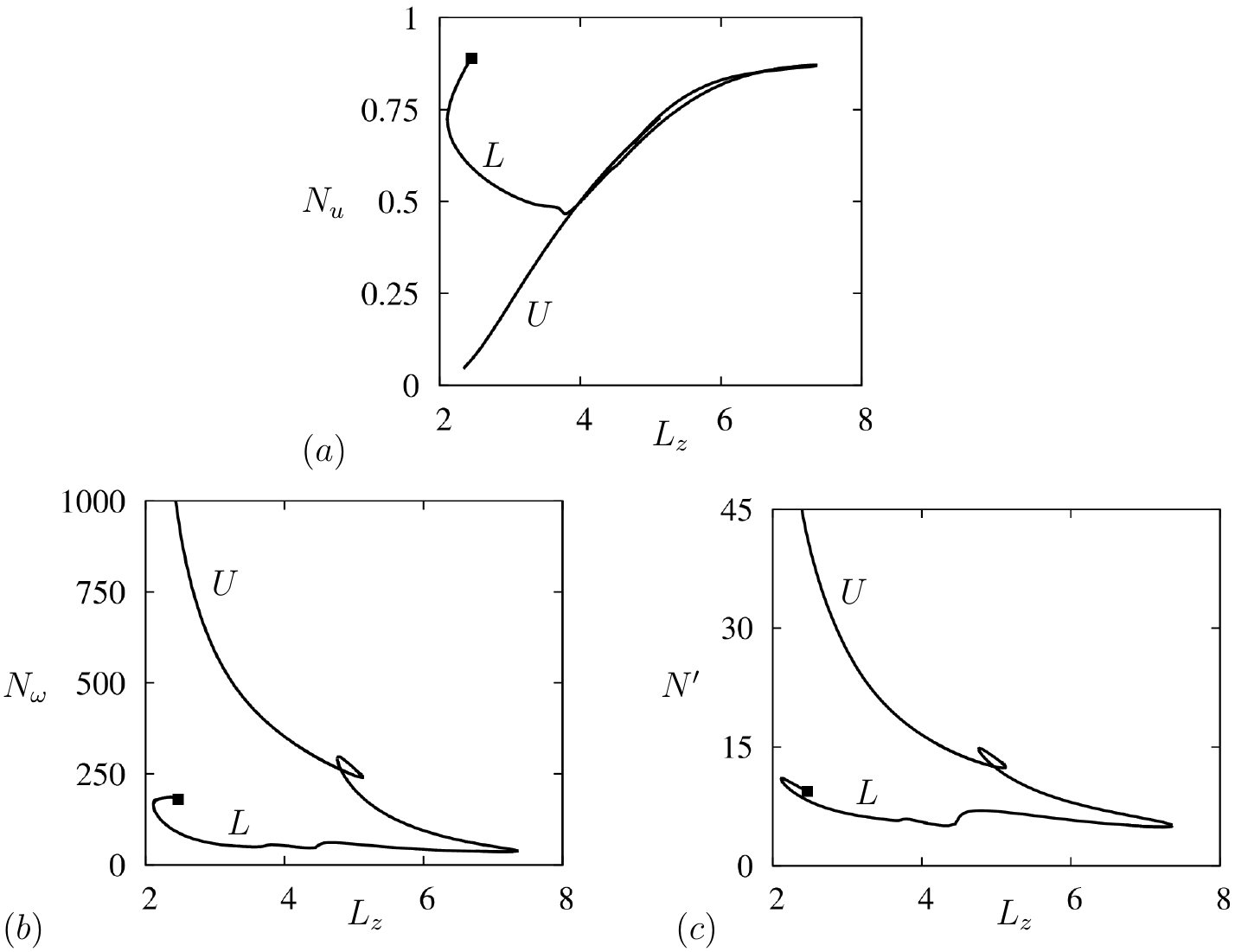}}
\caption{The ECS at $Re \approx 1500$ as a function of the spanwise period $L_z$. The diagrams show (a) $N_u$, (b) $N_{\omega}$ (c) and $N'$. The full square indicates the termination of the branch on a branch of solutions with two wavelengths in the domain (figure \ref{lowerksnap}, upper panel). Solutions along the lower branch (denoted by $L$) are shown in figure \ref{lowerksnap}, while those on the upper branch (denoted by $U$) are shown in figure \ref{upperksnap}.}
\label{kdiag}
\end{figure}

\begin{figure}
\centerline{\includegraphics[]{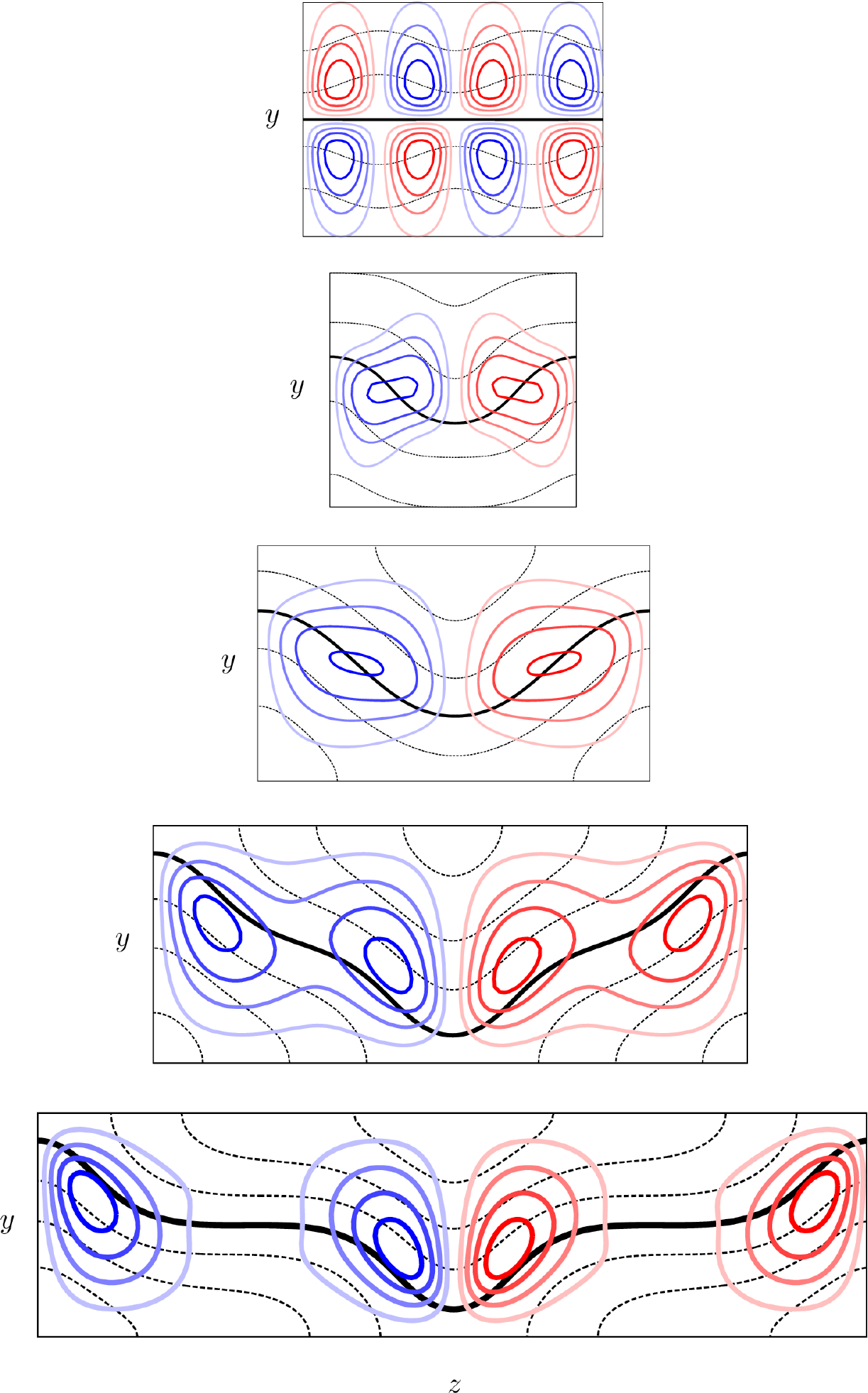}}
\caption{Structure of the lower branch solutions with different spanwise periods $L_z$, showing the contours of streamwise velocity in black (the solid line representing the critical layer) and contours of the mean streamfunction in color. From top to bottom: endpoint of the branch at $L_z \approx 2.5$, left saddle-node at $L_z \approx 2.1$, solution at $L_z = \pi$ (taken from figure \ref{lower}), solution at $L_z \approx 5$ and solution at the right saddle-node at $\L_z \approx 7.4$. The contour values of the streamwise velocity are the same throughout, but different values of the streamfunction are used from panel to panel for better representation of the flow. The contours are in all cases equidistributed.}
\label{lowerksnap}
\end{figure}

We next investigate how the solutions computed in the previous section depend on the spanwise domain size $L_z$ and the imposed streamwise wavenumber $\alpha = 2 \pi / L_x$. We begin by fixing $\alpha = 0.5$ and studying the effect of varying $L_z$. The resulting bifurcation diagrams are shown in figure \ref{kdiag}.
Continuation of the lower branch states to domains with smaller spanwise extent reveals that they pass a saddle-node at $L_z \approx 2.1$ before terminating on a branch of solutions with two wavelengths per period when $L_z \approx 2.5$. These solutions are depicted in the top two panels in figure \ref{lowerksnap} together with the initial lower branch solution at $L_z = \pi$ in the third panel.
The rolls present at $L_z = \pi$ gradually tilt as the domain period is reduced (figure \ref{lowerksnap}, second panel). Beyond the saddle-node at $L_z\approx 2.1$ these tilted vortices continue to stretch diagonally, in a direction transverse to the critical layer. As this happens the centre of each roll gradually splits forming two co-rotating rolls, one on either side of the critical layer, a process that greatly reduces the deflection of the critical layer from its laminar location at $y=0$. At the same time small counter-rotating rolls appear in the corners above and below each tilted structure and these grow in strength as $L_z$ increases, ultimately forming a period two state at $L_z\approx 2.5$ with an unperturbed $y=0$ critical layer (figure \ref{lowerksnap}, top panel). The resulting period-doubled ECS bears a number of similarities with the solutions EQ7 and EQ8 first observed by \cite{Gibson08} and reported in figure 16 of \cite{Gibson14}.

\begin{figure}
\centerline{\includegraphics[]{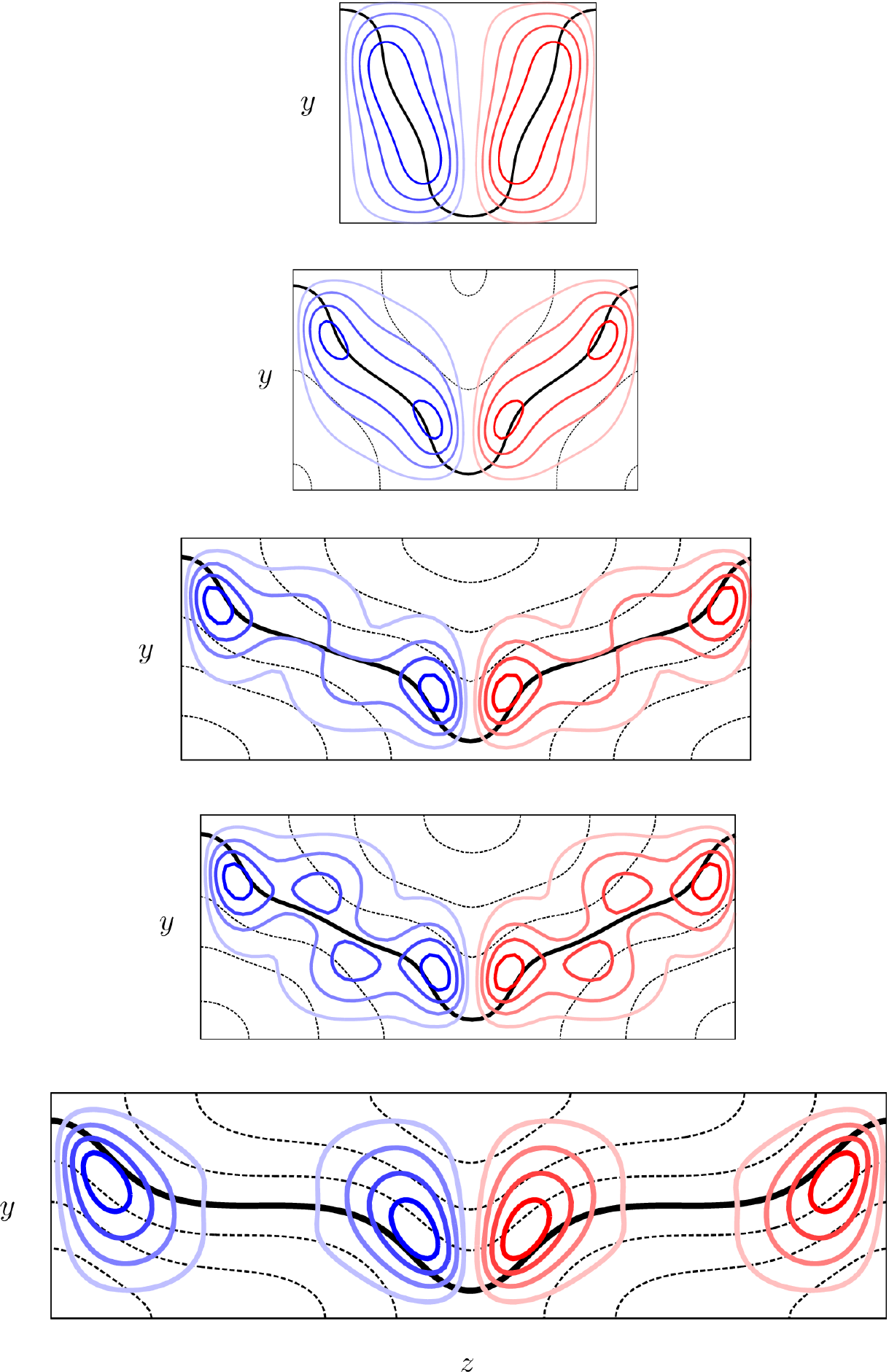}}
\caption{Structure of the upper branch solutions with different spanwise periods $L_z$, showing the contours of streamwise velocity in black (the solid line representing the critical layer) and contours of the mean streamfunction in color. From top to bottom: diverging solution at $L_z \approx 2.3$, solution at $L_z = \pi$ (taken from figure \ref{upper}), solution at the right saddle-node of the loop at $L_z \approx 5.1$, solution at the left saddle-node of the loop at $L_z \approx 4.8$ and solution at the right saddle-node at $\L_z \approx 7.4$ (taken from figure \ref{lowerksnap}). The contour values of the streamwise velocity are the same throughout, but different values of the streamfunction are used from panel to panel for better representation of the flow. The contours are in all cases equidistributed.}
\label{upperksnap}
\end{figure}

Continuing the lower branch state at $L_z=\pi$ in the opposite direction, towards larger $L_z$, reveals a new type of behavior. The increasing domain size stretches the rolls, which evolve into a bimodal structure reminiscent of the upper branch solution with $L_z=\pi$ (compare figure \ref{lowerksnap}, fourth panel, with figure \ref{upper}(a)). Increasing $L_z$ further leads to the progressive breakup of each of the original rolls into a pair of co-rotating rolls (figure \ref{lowerksnap}, fourth panel). Once formed these rolls are pulled farther apart as $L_z$ increases, resulting in a periodic array of pairs of counter-rotating rolls supporting a highly deformed critical layer interspersed with connecting zones where the trivial laminar flow is only weakly perturbed (figure \ref{lowerksnap}, bottom panel). The resulting state cannot be continued to larger domain sizes and passes a saddle-node at $L_z \approx 7.4$ where it connects with states originating along the upper branch (see below and figure \ref{upperksnap}). This type of behavior is similar to that observed for PCF by \cite{Deguchi13} but is not related to spatial localization in the spanwise direction as conventionally understood, since true localized states must become independent of the domain size. 

From the saddle-node solution (last panels in figures \ref{lowerksnap} and \ref{upperksnap}) one can continue the branch back to lower values of $L_z$ but in the direction of increasing fluctuation intensity, i.e., along the upper branch (see figure \ref{kdiag}(c)). The resulting upper branch states are shown in figure \ref{upperksnap}. As the period $L_z$ is reduced from $L_z \approx 7.4$ along the upper branch the stretching gradually disappears, but the co-rotating rolls do not merge, in contrast to the behavior along the lower branch. Instead, two additional rolls are nucleated between the orginal pair of co-rotating rolls, and these also co-rotate (figure \ref{upperksnap}, fourth panel). The net result is an array of four co-rotating rolls whose combined action deforms the critical layer further from the laminar case, and these are paired with a similar set of four co-rotating rolls in the other half of the domain, but rotating in the opposite sense. This four-roll structure is destroyed as the branch passes through a loop between $L_z \approx 4.8$ and $L_z \approx 5.1$: the two weaker middle rolls that have appeared along the upper branch below $L_z \approx 7.4$ gradually fade, thereby restoring the bimodal structure (figure \ref{upperksnap}, second panel) and generating the state in figure \ref{upper}. When $L_z$ is decreased further, the bimodal structure gradually disappears as the rolls are squeezed together (figure \ref{upperksnap}, top panel). At the same time, both the mean streamfunction and the fluctuation fields grow without bound while the streamwise velocity $u_0$ becomes increasingly homogenized and ultimately approaches zero. These developments are reflected in the dramatic decrease in the streamwise velocity norm $N_u$, together with increases in enstrophy norm $N_{\omega}$ and fluctuation norm $N'$ shown in figure \ref{kdiag}, and indicate that the postulated asymptotic form of the solutions is starting to break down. Thus solutions in this regime are unlikely to be physically relevant.

We mention that in related calculations for PCF, \cite{Melnikov14} find that the lower and upper branch states form an isola in $L_z$ with no additional bifurcations (except for saddle-nodes) as $L_z$ varies. 

\begin{figure}
\centerline{\includegraphics[]{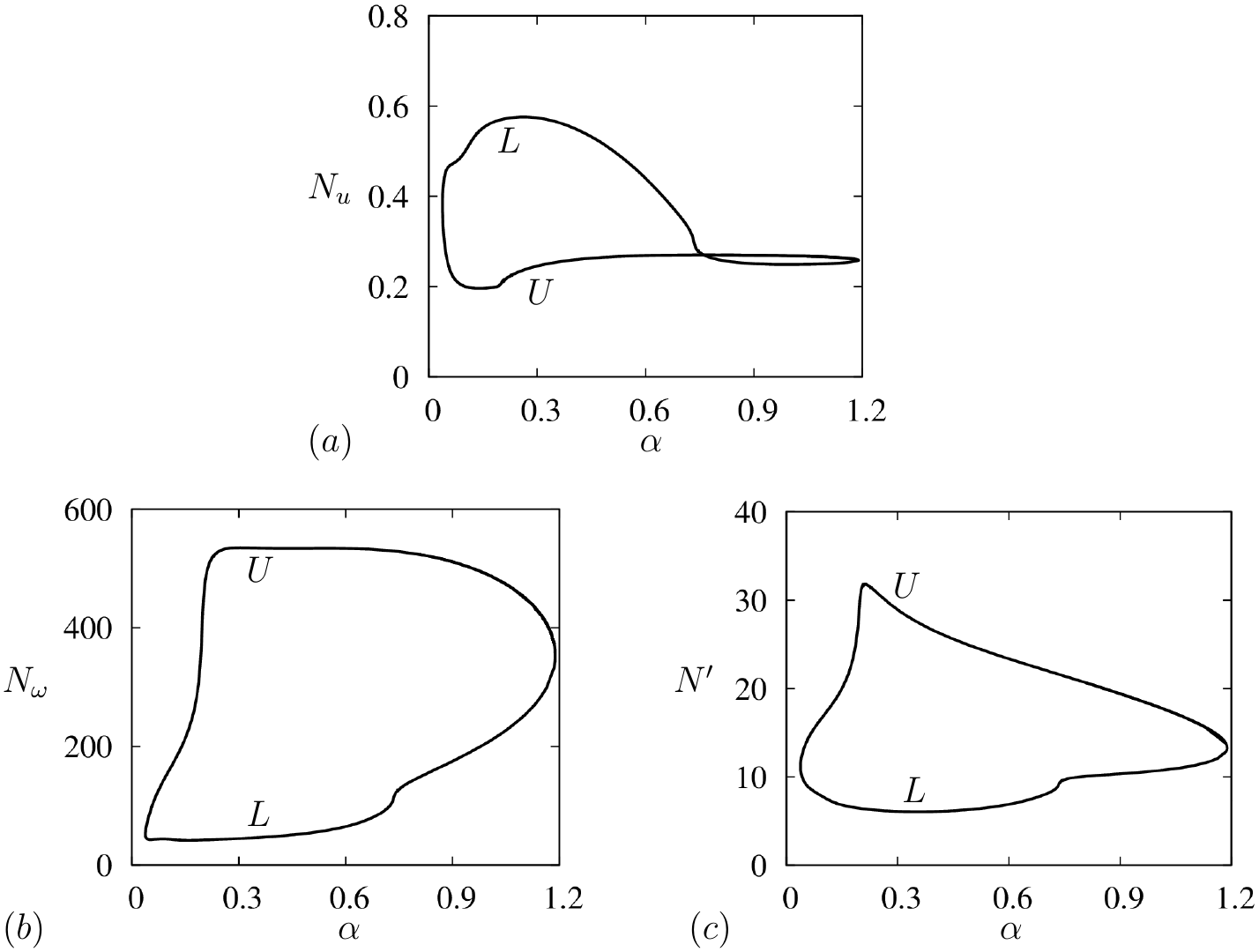}}
\caption{Bifurcation diagrams showing the ECS at $Re = 1500$ and $L_z = \pi$ as a function of the streamwise wavenumber $\alpha$. (a) $N_u$. (b) $N_{\omega}$. (c) $N'$. Solutions at the left and right saddle-nodes are shown in figure \ref{lowerasaddle}. The letter $L$ (resp. $U$) denotes the lower (resp. upper) branch.}
\label{adiag}
\end{figure}

\begin{figure}
\centerline{\includegraphics[]{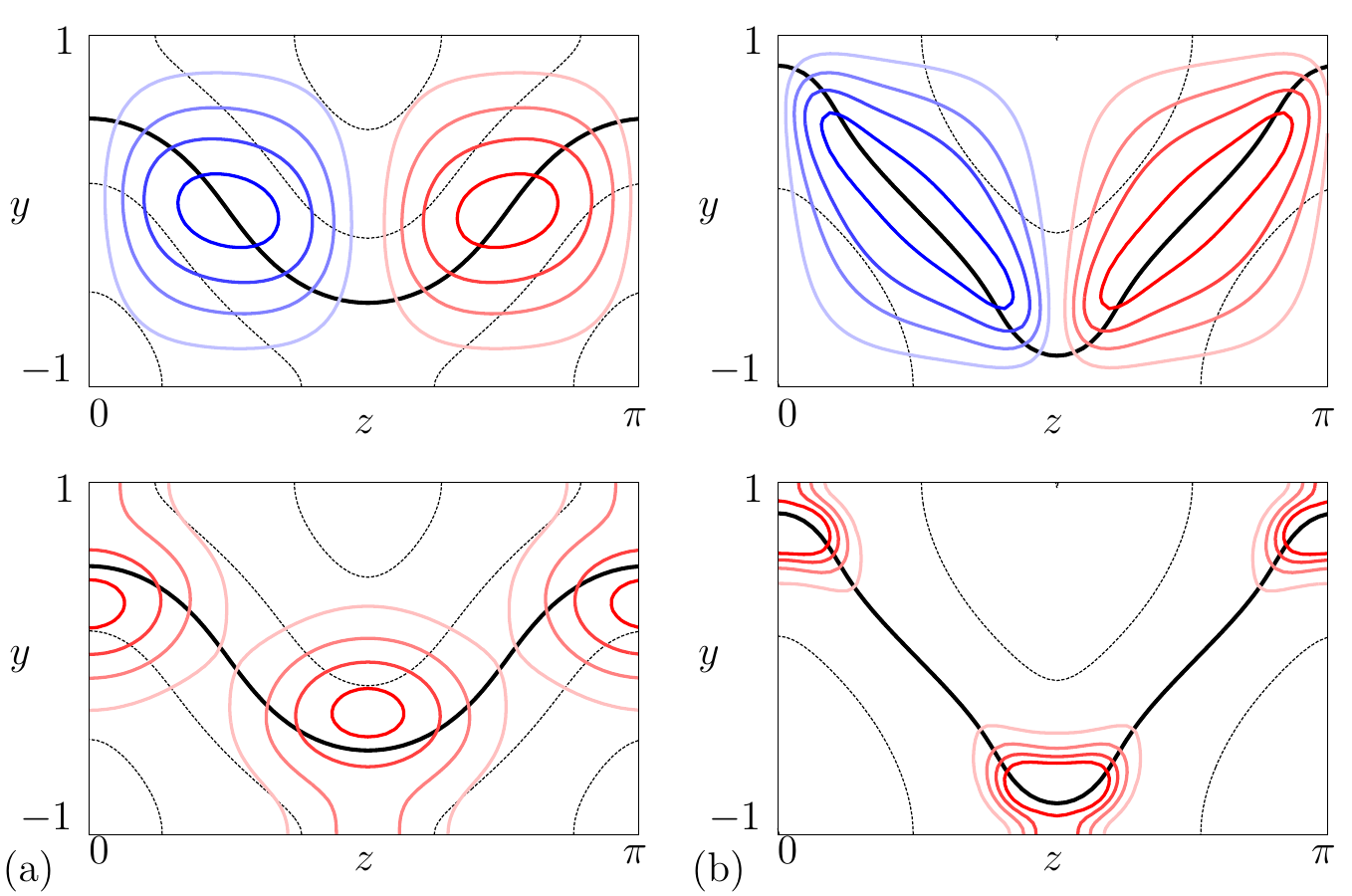}}
\caption{Structure of the ECS at (a) the left saddle-node at $\alpha \approx 0.0380$ ($L_x \approx 165$) and (b) the right saddle-node at $\alpha \approx 1.1890$ ($L_x \approx 5.3$) of the isola shown in figure \ref{adiag}. The solutions are represented using contours of constant mean streamwise velocity in black (the solid line representing the critical layer) with superposed contours of the streamfunction in color (top panel) and the fluctuation norm (bottom panel). The contours are in all cases equidistributed.}
\label{lowerasaddle}
\end{figure}

We have also studied how the morphology of the ECS in a domain with $L_z = \pi$ varies at fixed $Re = 1500$ when the streamwise wavenumber $\alpha$ is changed. The bifurcation diagrams obtained are shown in figure \ref{adiag}.
The figure reveals that the ECS lie on an isola and therefore do not connect to any other solution. Thus the isola defines an interval of existence for the ECS at $L_z = \pi$ and $Re = 1500$: $0.0380 < \alpha < 1.1890$. While the lower bound for $\alpha$ hints at the persistence of these structures for very long domains ($L_x \approx 165$), the upper bound indicates that the required streamwise periodicity of the domain be at least $L_x \approx 5.3$ for these structures to be self-sustaining, a value close to that observed in PCF \citep{Hall10}. The ECS at the left and right saddle-nodes along the isola are represented in figure \ref{lowerasaddle}. 
For small $\alpha$ (long streamwise domain), the critical layer is distorted approximately sinusoidally by nearly circular rolls and the fluctuations do not exhibit sharp gradients despite being located close to the critical layer $u_0 = 0$. In contrast, for large $\alpha$ and therefore short streamwise domains, the fluctuations become very localized and the associated critical layer is deformed into a sawtooth profile. The associated rolls are highly elongated and align with the approximately constant slope sections of the critical layer. These developments are a consequence of the $(\alpha Re)^{-1/3}$ critical layer scaling \citep{Maslowe86}. As $Re$ is kept fixed, decreasing $\alpha$ increases the width of the critical layer thereby weakening the strength of the fluctuations. This is a direct consequence of the fact that $\alpha$ affects the amplitude of the $u_0$-induced advection of fluctuations, as described by Eq.~(\ref{rdc4}). As a result decreasing the value of $\alpha$ decreases the coupling between the mean and fluctuation fields.  Thus, the fluctuations no longer track the critical layer efficiently, and so remain weak and do not deform the rolls. The associated critical layer is sinusoidal. The converse is true when $\alpha$ is increased: the coupling becomes stronger, leading to fluctuations that are strongly focused on the critical layer and that substantially deform the rolls and hence generate a strongly distorted critical layer. 

\section{Discussion}\label{Discussion}

In this paper we have presented a simple asymptotic procedure that leads to a reduced description of plane parallel shear flows. The method assumes that the flow is dominated by the mean streamwise flow component, with spanwise components of the velocity field (the rolls) that are, in the limit $Re\to\infty$, much weaker. Despite this the Reynolds stress generated by the fluctuating fields modifies the mean spanwise velocity and hence the mean streamwise flow, as described by Eqs.~(\ref{U0barEQN})--(\ref{CONTbarEQN}), and this in turn modifies the fluctuations as described by Eqs.~(\ref{U1primeEQN})--(\ref{CONTprimeEQN}). Of these the former are simplified in having an $O(1)$ effective Reynolds number while the latter constitute a singular but quasilinear system that admits solutions of arbitrary amplitude. In our approach this amplitude is determined by a self-consistency requirement: in steady state the Reynolds stress generated by the fluctuations must be such as to produce a streamwise flow for which the fluctuations neither grow nor decay, and we have described an iterative process whereby the amplitude of the fluctuations can be adjusted to realize this requirement.  Our approach therefore captures the essence of the self-sustaining mechanism identified by \cite{Waleffe97}.

In our reduced model we retain subdominant viscous terms to regularize the critical layer (cf.~\cite{BeaumeGFD12,Blackburn13}), thereby reintroducing a parameter that we profitably use for numerical continuation; in the present case this parameter is naturally identified with the inverse Reynolds number, $\epsilon\equiv 1/Re$, but we emphasize that it is fundamentally a homotopy parameter that can be used to identify different types of solutions at large $Re$.  While solutions at finite $Re$ obtained by this procedure cannot be exact, the properties of such solutions appear to be captured qualitatively correctly, and it is in this sense that our approach may prove to be particularly useful. In fact, we believe that the reduced equations capture the universality in the behavior of plane parallel shear flows and are suitable not only for studying steady ECS with critical layers but also exact traveling waves and indeed other nonequilibrium structures with no critical layer at all. In this respect the equations possess advantage over detailed studies of particular flows using flow-specific scalings. However, to justify these claims our results on WF and other flows will have to be compared quantitatively with solutions of the corresponding fully three-dimensional problems. Such comparisons \citep{Beaume_in_prep} will, in addition, determine, on a case by case basis, the range of Reynolds numbers for which our results provide a reliable guide to the solutions of the full problem.

We have applied the algorithm to compute a variety of exact coherent states in a body-force driven flow we refer to as Waleffe flow. The fundamental assumptions we make turns out to capture not only the expected lower branch states but also the corresponding upper branch states, reached via numerical continuation in the parameter $\epsilon$. 
The results we obtain are similar to the corresponding PCF results obtained by \cite{Blackburn13} for lower branch states and by \cite{Deguchi14} for upper branch states. In both systems the lower branch critical layer deforms into a sinusoidal surface through the action of the rolls, while the deformation corresponding to the upper branch states is both stronger and bimodal. Particularly intriguing is our discovery that along the upper branch the width of the critical layer is no longer uniform and that the bimodal structure of the rolls and streaks concentrates the critical layer forcing in regions of maximum departure from the unperturbed critical layer. Current asymptotic approaches do not take this possibility into account. However, the intrinsic self-consistency of our reduced equations implies that this new critical layer structure is likely a property of upper branch states in the full system at large $Re$.

We have used numerical continuation to continue our solutions in the spanwise domain length $L_z$ and in the streamwise wavenumber $\alpha$. The former determines the existence region for the solutions we have found and shows that the lower branch solutions at small $L_z$ bifurcate from a period two spatially periodic state but undergo a saddle-node bifucation at larger $L_z$ that connects the lower and upper branch states. The solutions near this fold are stretched in the spanwise direction relative to $O(1)$ domains but are not spatially localized in the conventional sense, in contrast to the suggestion made by \cite{Deguchi13}, since they cannot be continued to larger $L_z$ and hence to larger separations. The continuation in the streamwise wavenumber $\alpha$ leads to simpler results -- this time the lower and upper branch states are connected by folds at either end and the solutions lie on an isola. As a result they do not extend to either very small or very large values of $\alpha$.

Despite our success in computing exact coherent structures in Waleffe flow as a function of both $L_z$ and $\alpha$, at large values of $Re$, all the solutions computed appear to be unstable. Therefore, direct numerical simulation of the reduced equation set does not result in solutions that can be compared with direct numerical simulations of the primitive equations -- at least for Waleffe flow in the parameter regime explored -- in contrast to similar reductions for convection in a strong magnetic field \citep{JK07} or rapidly rotating convection \citep{JKRV13,RJKW14}. Despite this drawback we believe that the states we have computed will be of great value in further explorations of the full set of reduced equations, including studies of spatial modulation and possible localization in the streamwise direction. We hope to report on these explorations in a future publication.


{\bf Acknowledgement:}
This work was initiated in 2009 and the first numerical results were obtained as part of a summer research project of C. Beaume at the 2012 Geophysical Fluid Dynamics Program at the Woods Hole Oceanographic Institution \citep{BeaumeGFD12}. The work was supported by the National Science Foundation under grants DMS-1211953 (CB \& EK), OCE-0934827 (GPC) and OCE-0934737 (KJ). E.K. wishes to acknowledge additional support from the Chaire d'Excellence Pierre de Fermat de la r\'egion Midi-Pyr\'en\'ees (France).




\bibliographystyle{jfm}

\bibliography{wally}

\end{document}